\documentclass[journal]{IEEEtran}

\usepackage{color}
\usepackage{amssymb}
\usepackage{amsmath}
\usepackage{amsfonts}
\usepackage{graphicx}
\usepackage{epstopdf}
\usepackage{epsf, epsfig, textcomp}
\usepackage{dsfont}
\usepackage{balance}
\usepackage{amsthm}
\usepackage{amsthm}
\usepackage{algpseudocode}
\usepackage{algorithm}

\begin{document}
\title{Limited Feedback  Channel Estimation in Massive MIMO with Non-uniform Directional Dictionaries}

\author{Panos~N.~Alevizos,~\IEEEmembership{Student~Member,~IEEE,}
Xiao Fu,~\IEEEmembership{Member,~IEEE,}
Nicholas D. Sidiropoulos,~\IEEEmembership{Fellow,~IEEE,} \\
Ye Yang,~\IEEEmembership{}and~Aggelos~Bletsas,~\IEEEmembership{Senior~Member,~IEEE}%
\thanks{P.~N.~Alevizos and A. Bletsas are with   School of Electrical
and Computer Engineering, Technical University of Crete, Chania 73100, Greece
(e-mail: palevizos$@$isc.tuc.gr; aggelos$@$telecom.tuc.gr).

X. Fu is with the School of Electrical Engineering and Computer Science, Oregon State University (e-mail: xiao.fu$@$oregonstate.edu). 

N. D. Sidiropoulos is with the Department of Electrical and Computer Engineering, University of Virginia, Charlottesville, VA 22904 (e-mail: nikos$@$virginia.edu).

Ye Yang is with Physical Layer $\&$ RRM IC Algorithm Dept., WN Huawei Co., Ltd. (e-mail:  yangye$@$huawei.com).}
 }

\maketitle

\IEEEpeerreviewmaketitle
%cite{LeeHanZh:09, SeTaBa:11} \cite{LTE_transmi_modes:15}
%Changed by evlachos in v_5

\begin{abstract}
Channel state information (CSI) at  the base station (BS) is crucial to achieve beamforming and multiplexing gains in multiple-input multiple-output 
(MIMO) systems. State-of-the-art limited feedback schemes require feedback overhead that scales linearly with the number of BS antennas, which 
is prohibitive for $5$G massive MIMO. This work proposes  novel limited feedback algorithms that lift this burden by exploiting the inherent 
sparsity in double directional (DD) MIMO channel representation  using overcomplete dictionaries. These dictionaries are associated with angle of
arrival (AoA) and angle of departure (AoD) that specifically account for antenna directivity patterns at both ends of the link. The proposed 
algorithms achieve satisfactory channel estimation accuracy using a small number of feedback bits, even when the number of transmit antennas 
at the BS is large -- making them ideal for $5$G massive MIMO. Judicious simulations reveal that they outperform a number of popular feedback
schemes, and underscore the importance of using angle dictionaries matching the given antenna directivity patterns, as opposed to uniform dictionaries.
The proposed algorithms are   lightweight  in terms of computation, especially on the user equipment side, making them ideal for actual deployment in $5$G systems.
\end{abstract}


\section{Introduction}
\label{sec:intro}

The idea of harnessing a large number of antennas at the base station (BS),  possibly many more than the number of user equipment (UE) terminals
in the cell, has recently attracted a lot of interest in  massive  multiple-input multiple-output (MIMO) research.
The key technical reasons for this is that massive MIMO can enable leaps in spectral efficiency \cite{HoyBriDeb:13}
as well as help mitigating intercell interference through  simple linear precoding and combining, offering immunity to
small-scale fading -- known as the channel hardening effect  \cite{Mar:10, RuPeLauLarMaeEdTuf:13}. Massive MIMO
systems also have the advantage of being energy-efficient since every antenna may operate at a low-energy level \cite{NgLarMar:13}.

Acquiring accurate and timely downlink channel state information (CSI) at the BS is the key to realize
the multiplexing and array gains enabled by MIMO systems \cite{Mar:10, Jin:06, CaJiKoRa:10,AdNamAhnCai:13}. Acquiring accurate
downlink CSI at the BS using only few feedback bits from the UE is a major challenge, especially in massive MIMO systems. 
In frequency division duplex (FDD) systems, where channel reciprocity does not hold, the BS cannot acquire downlink channel
information from uplink training sequences, and the feedback overhead may be required to scale proportionally to the number of BS antennas \cite{Jin:06}.
In time division duplex (TDD) systems, channel reciprocity between uplink and downlink is often assumed, and the BS acquires
downlink CSI through uplink training. Even in TDD mode, however, relying only on channel reciprocity is not accurate enough,
since the uplink measurements at the BS cannot capture the downlink interference from neighboring cells  \cite{DahParkSk:16,Hyo_et_all:16}. 
Thus, downlink reference signals are still required to estimate and feed back the channel quality indicator (CQI), meaning that some 
level of feedback is practically necessary for  both FDD and TDD modes.

The largest portion of the feedback-based channel estimation literature explores various quantization techniques; see \cite{Love_et_al:08}
for a well-rounded exposition. Many of these methods utilize a vector quantization (VQ) codebook that is known to both the BS and the UE. 
After estimating the instantaneous downlink CSI at the UE, the UE sends through a limited feedback channel the index of the codeword that best matches
the estimated channel, in the sense of minimizing the outage probability \cite{MuSabErkAaz:03},  maximizing link capacity \cite{LaLiCh:04}, or
maximizing the beamforming gain \cite{RyClVauCoGuoHon:09, ChoLovMad:13}.
Codebooks for spatially correlated channels based on generalizations of the Lloyd algorithm are given in \cite{XiaGian:06},
while codebooks  designed for temporally correlated channels are provided  in \cite{HuHeAn:09}. 
Codebook-free feedback for channel tracking was considered in \cite{MehSid:14_a} for spatio-temporally correlated
channels with imperfect CSI at the UE. Many limited feedback approaches in MIMO systems consider a Rayleigh fading channel 
model  \cite{MarHoch:06, RyClVauCoGuoHon:09, ChoLovMad:13, JiaMoCaiZhi:15}.
Under this channel model, the number of VQ feedback bits required to guarantee reasonable performance is linear in the number of transmit antennas at the BS 
\cite{Jin:06} -- which is costly in the case of massive MIMO. Yet the designer is not limited to using VQ-based approaches, and massive MIMO channels can be far from Rayleigh.

In this work, we consider an approach that differs quite sharply from the prevailing limited feedback methodologies.
Our approach specifically targets FDD massive MIMO in the sublinear feedback regime. We adopt the double directional (DD) MIMO channel model
\cite{BaHaSaNo:10} (see also \cite{HeGoRaRoSay:16}) instead of the Rayleigh fading model. 
The DD channel model parameterizes each channel path using angle of departure (AoD) at BS, small- and large-scale
propagation coefficients, and angle of arrival (AoA) at UE -- a parametrization that is well-accepted and advocated 
by 3GPP \cite{3GPP_TS_36101_Rel13, KamKhAltDebKam:14}. We exploit a `virtual sparse representation' of the downlink channel
under the double directional MIMO model \cite{BaHaSaNo:10}. Quantizing AoA and AoD, it is possible to design overcomplete
dictionaries that contain steering vectors approximating those associated with the true angles of arrival and departure. 
Building upon \cite{BaHaSaNo:10}, such representation has been exploited to design receiver-side millimeter wave (mmWave) channel estimation algorithms 
using high-resolution \cite{MerRusGonAlkHe:16}, or low-resolution (coarsely quantized) analog-to-digital converters (ADCs) \cite{MoShPrHe:14, RuRiPrHe:15}.

In contrast, we focus on transmitter-side (BS) downlink channel acquisition using only limited receiver-side (UE) computation and
feedback to the BS \cite{ZhChGuHo:17}. We propose novel optimization formulations and algorithms for downlink channel estimation at the 
BS using single-bit judiciously-compressed measurements. In this way, we shift the channel computation burden from the UE 
to the BS, while keeping the feedback overhead low. Using the overcomplete parametrization of the DD model, three new limited feedback setups  are proposed:
\begin{itemize}
\item  In the first setup,   UE applies   dictionary-based sparse channel estimation and support identification 
to estimate the 2D angular support and the corresponding coefficients of the sparse channel. 
Then, the UE feeds back the support of the sparse channel estimate, plus a coarsely quantized version  of the  corresponding non-zero coefficients,
assuming known thresholds at the BS.
This is the proposed {\em UE-based limited feedback baseline} method for the DD model.
\item
In the second setup, the UE  compresses the received measurements and sends back only the signs of the compressed measurements to the BS. Upon receiving these sign bits, the BS estimates the channel using {\em single-bit DD dictionary-based sparse estimation} algorithms.
\item The third setup is a combination of the first and the second, called {\em hybrid limited feedback}: UE estimates and sends the support of the
sparse channel estimate on top of the compressed sign feedback used in the second setup. Upon receiving this augmented feedback from the UE, the BS can then apply the algorithms of setup 2 on a significantly reduced problem dimension.
\end{itemize}
For sparse estimation and support identification, the orthogonal matching pursuit (OMP) algorithm \cite{TroGil:07} is utilized as it offers the best possible computational complexity among all sparse estimation algorithms \cite{ChDiRaKi:15}, which is highly desired for resource-constrained UE terminals.

\smallskip
\noindent
{\bf Contributions:}

A new limited feedback channel estimation framework is proposed exploiting the sparse nature of the DD model (setup 2).
Two formulations are proposed based on single-bit sparse maximum-likelihood estimation (MLE) and single-bit compressed sensing. 
For MLE,  an optimal in terms of iteration complexity \cite{Nes:04} first-order proximal  method is designed using  adaptive restart, 
to further speed up the convergence rate   \cite{ODonCan:15}. The proposed compressed sensing (CS) formulation can 
be -- fortuitously -- harnessed by invoking the recent single-bit CS literature. The underlying convex optimization problem
has a simple closed-form solution, which is ideal for practical implementation. The proposed framework shifts the computational
burden towards the BS side -- the UE only carries out matrix-vector multiplications and takes signs. This is sharply different 
from most limited feedback schemes in the literature, where the UE does the `heavy lifting' \cite{CaJiKoRa:10, Love_et_al:08}. 
More importantly, under our design, using a small number of feedback bits achieves very satisfactory channel estimation accuracy even when the
number of  BS antennas is very large, as long as the number of paths is reasonably small -- which is usually the case in practice \cite{BaHaSaNo:10};
thus, the proposed framework  is ideal for massive MIMO 5G cellular networks.

In addition to the above contributions, a new angle dictionary construction methodology is proposed to enhance performance, 
based on a companding quantization technique \cite{GrNeu:99}.
The idea is  to create dictionaries that concentrate the angle density in a non-uniform manner, around the angles where 
directivity patterns attain higher values. The baseline 3GPP  antenna directivity pattern is considered for this, and the 
end-to-end results are contrasted with those obtained using uniform quantization, to showcase this important point.
Judicious simulations reveal that the proposed dictionaries   outperform uniform dictionaries.

Last but not least, to further reduce computational complexity at the BS and  enhance beamforming and ergodic rate performance, 
a new hybrid implementation is proposed (setup 3). This setup
is very effective when the UE is capable of carrying out simple estimation algorithms, such as OMP. 
At the relatively small cost of communicating extra support information that slightly increases feedback
communication overhead, the BS applies the single-bit MLE and single-bit CS algorithms
on a dramatically reduced problem dimension. Simulations reveal that
the performance of the two algorithms under setup 3 is always better than under setup 2. As in setup 2, 
the feedback overhead is tightly controlled by the system designer and the desired level of channel estimation accuracy is
attained with very small feedback rate, even in the massive MIMO regime.

Comprehensive simulations over a range of pragmatic scenarios,  based on the 3GPP DD  channel model \cite{3GPP_TR_36814_Rel9},
compare the proposed methods with baseline least-squares (LS) scalar and
vector quantization (VQ) feedback strategies in terms of normalized mean-squared estimation error (NRMSE), beamforming gain,
and  multi-user capacity under zero-forcing (ZF) beamforming. Unlike VQ, which requires that the number of feedback bits grows at least linearly with the number of BS antennas
to maintain a certain level of estimation performance, the number of feedback bits of the proposed algorithms is controlled
by the system designer, and substantial feedback overhead reduction is observed for achieving  better performance compared to VQ methods.
It is also shown that when the sparse DD model is valid, the proposed methods not only outperform LS schemes, but they may also offer
performance very close to perfect CSI in some cases.

Relative to the conference precursor \cite{AlFuSidYaBl:17} of this work, this journal version
includes the following additional contributions: the UE-based limited feedback scheme under
setup 1; the novel channel estimation algorithm based on the sparse MLE
formulation; %derivation of the closed-form CS solution;
the new hybrid schemes under setup 3;
and comprehensive (vs. illustrative) simulations of all schemes considered.
The rest of this paper is organized as follows.
Section~\ref{sec:sys_model} presents the adopted wireless system model, and 
Section~\ref{sec:angle_dict_construction} derives the proposed
non-uniform directional dictionaries.
Sections~\ref{sec:UE_based_limited_feedback_sparse_channel_estimation},~\ref{sec:Tx_based_limited_feedback_sparse_channel_estimation},
and~\ref{sec:hybrid_limited_feedback_sparse_channel_estimation} develop the proposed UE-based, BS-based, and hybrid limited feedback algorithms, respectively.
Section~\ref{sec:numerical_results} presents simulation results, and 
Section~\ref{sec:conclusion} summarizes conclusions. 

\emph{Notation}:
Boldface lowercase and uppercase letters denote  column vectors and matrices,
respectively;
 $()^*$, $()^{\top}$,  and $()^{\mathsf{H}}$, denote conjugate,
transpose, and Hermitian operators, respectively.
$\| \cdot \|_p$, $\Re(\cdot)$, $\Im(\cdot)$, and $|\cdot|$
denote the $p$-norm (with $p \in [0,\infty]$), the real, the imaginary, and the
 absolute or set cardinality operator,  respectively.
$\mathsf{diag}(\mathbf{x})$ is the diagonal matrix formed by vector $\mathbf{x}$, $\mathbf{0}$ is the all-zero
vector and its size is understood from the context, $\mathbf{I}_N$ is the $N\times N$
identity matrix.
% Symbols $\nabla \mathsf{f}(\cdot)$ and $\nabla^2 \mathsf{f}(\cdot)$ are the gradient and the Heassian matrix of function $\mathsf{f}$.
Symbol $\otimes$ denotes  the Kronecker product.
 $\mathbb{E}[\cdot]$ is the expectation operator.   $\mathcal{CN}(\boldsymbol{\mu}, \boldsymbol{\Sigma})$
denotes the proper complex Gaussian distribution with mean $\boldsymbol{\mu}$ and covariance $ \boldsymbol{\Sigma}$.
Matrix (vector) $\mathbf{A}_{:,\mathcal{S}}$ ($\mathbf{x}_{\mathcal{S}}$) comprises of the columns of matrix  $\mathbf{A}$ (elements of
$\mathbf{x}$)
indexed by set $\mathcal{S}$.
Function $\mathsf{sign}(x)=1$ for $x\geq0$ and zero, otherwise; abusing notation a bit, we also apply it to vectors, element-wise. Function $(x)_+ = \max(0, x) $,
$\mathsf{j} \triangleq \sqrt{-1}$ is the imaginary unit,  and  $\mathsf{Q}(x) =
\frac{1}{\sqrt{2 \pi}}\int_{x}^{\infty} \mathsf{e}^{-t^2 /2} \mathsf{d} t$ is the $\mathsf{Q}$-function.
%  $\partial \mathsf{f}(\mathbf{x})$ is the subdifferential of   function $\mathsf{f}$ given by $\partial \mathsf{f}(\mathbf{x}) = \{ \mathbf{q} : \mathsf{f}(\mathbf{y}) \geq
% \mathsf{f}(\mathbf{x}) + \mathbf{q}^{\top} (\mathbf{y} - \mathbf{x}),\, \forall \mathbf{y} \in \mathbf{dom}\mathsf{f}\}$.

\section{System Model}
\label{sec:sys_model}

We consider an FDD cellular system consisting of a BS serving $K$ active UE terminals, where the downlink channel is estimated at the BS through feedback from each UE. For brevity of exposition, we focus on a single UE.
  The proposed algorithms can be easily generalized to  multiple users,
as the downlink channel estimation process can be performed separately for each UE. 
The BS is equipped with $M_{\rm T}$ antennas  and the UE is equipped with $M_{\rm R}$ antennas.
The channel is assumed static over a coherence block of $U_{\rm c} = B_{\rm c} T_{\rm c}(\frac{T_{\rm s}-T_{\rm g}}{T_{\rm s}})$
 complex orthogonal frequency division multiplexing (OFDM) symbols, where
 $B_{\rm c}$ is the coherence bandwidth (in Hz),  $T_{\rm c}$ is the coherence time (in seconds),
and quantity $(\frac{T_{\rm s}-T_{\rm g}}{T_{\rm s}})$ indicates the fraction of useful symbol time
(i.e., $T_{\rm s}$ is the OFDM symbol duration and $T_{\rm g}$ is the cyclic prefix duration).%
\footnote{
In LTE,   time-frequency resources   are structured in a such a way, so
 the coherence block occupies some  resource blocks -- each resource block   consists of $7$ contiguous OFDM symbols in time
 multiplied  by $12$ contiguous subcarriers in frequency.
A  subframe of duration $1$ msec consists of two contiguous in time resource blocks, yielding $2 \cdot 84 = 168$ symbols,
over which  the channel can be considered constant \cite{SeTaBa:11}.}
In downlink transmission, the BS has to acquire
CSI through feedback from the active UE terminals, and then
design the transmit signals accordingly. At the training phase,
the BS employs $N_{\rm tr}$ training symbols for channel estimation.
%a  baseband  precoder $\mathbf{W} \in \mathds{C}^{M_{\rm T} \times M_{\rm T}}$ and
%transmits the signal the
The narrowband  (over time-frequency)  discrete  model  over a period of $N_{\rm tr}$ training symbols is given by
\begin{equation}
\mathbf{y}_n = \mathbf{H}     \,  \mathbf{s}_n + \mathbf{n}_n, ~~n = 1, 2, \ldots, N_{\rm tr},
\label{eq:baseband_equivalent}
\end{equation}
where $n$ is the $n$-th training index, $\mathbf{s}_n  \in \mathds{C}^{M_{\rm T}}$ is the transmitted training signal,
  $\mathbf{y}_n \in \mathds{C}^{M_{\rm R}}$ is the
received vector,  $\mathbf{H} \in \mathds{C}^{M_{\rm R} \times M_{\rm T}}$ denotes  the complex baseband equivalent channel matrix, and
$\mathbf{n}_n \sim \mathcal{CN}(\mathbf{0} , \sigma^2 \mathbf{I}_{M_{\rm R}})$ is additive Gaussian noise
at the receiver of variance $\sigma^2$. All quantities in the right hand-side of~\eqref{eq:baseband_equivalent}
are independent of each other;  $\mathbb{E}[\mathbf{s}_n \mathbf{s}_n^{\mathsf{H}}] = \frac{P_{\rm T}}{M_{\rm T}}
 \mathbf{I}_{M_{\rm T}}$, for all $n$, where $P_{\rm T}$ denotes the average total transmit power.
The  signal-to-noise ratio (SNR)  is defined as
$
{\rm SNR} \triangleq \frac{ P_{\rm T}}{  \sigma^2} $.
% =\frac{P_{\rm T} \, \mathbb{E }[\|  \mathbf{H} \|_{F}^2] }{M_{\rm T}\,  M_{\rm R} \, \sigma^2}.
% \label{eq:avg_rcvd_SNR}
% \end{equation}

To estimate $\mathbf{H}$, we can use linear least-squares (LS) \cite{MarHoch:06}, or, if the
 channel covariance is known, the linear minimum mean-squared error (LMMSE)
approach \cite{CaJiKoRa:10}. These linear approaches need more than $M_{\rm T} M_{\rm R}$
training symbols to establish identifiability of the channel (to
`over-determine' the problem) -- which is rather costly in massive MIMO scenarios.

A more practical approach to the problem of downlink channel acquisition at the BS of
massive MIMO systems would be to shift the computational burden to the BS, relying on
relatively lightweight computations at the UE, and assuming that only low-rate feedback is available as well.
The motivation for this is clear: the BS is connected to the communication backbone,
plugged to the power grid, and may even have access to cloud computing -- thus is far
more capable of performing intensive computations. The challenge of course is how to
control the feedback overhead -- without a limitation on feedback rate, the UE can of
course simply relay the signals that receives back to the BS, but such an approach is clearly
wasteful and impractical. The ultimate goal is to achieve accurate channel estimation
with low feedback overhead, i.e., estimate $\mathbf{H}$ using just a few feedback bits.

Towards this end, our starting idea is to employ a finite scatterer (also known as discrete multipath, or double directional) channel model comprising of
$L$ paths, which can be parameterized using a virtual sparse representation.  
The  inherent  sparsity of DD parameterization in the angle-delay domain can be exploited also at the
the UE side to estimate   the downlink channel  using
compressed sensing techniques with  reduced  pilot sequence overhead \cite{BaHaSaNo:10}. This sparse
representation will lead to a feedback scheme that is rather parsimonious in terms of both overhead and computational complexity.
The narrowband downlink channel matrix $\mathbf{H}$ can be written as %
% \footnote{It is noted that the  narrowband  model can be easily extended to handle frequency-selectivity as well as the Doppler shifts,
% see \cite{BaHaSaNo:10} and references therein.}%
\begin{equation}
\mathbf{H} \!=\! \sqrt{\frac{M_{\rm T} M_{\rm R}}{L}} \sum_{l=1}^L \alpha_l  \,\mathsf{c}_{\rm T}(\phi_l' ) \, \mathsf{c}_{\rm R}(\phi_l )  \, \mathbf{a}_{\rm R }(\phi_l)
\, \mathbf{a}_{{\rm T}}^{\mathsf{H}}(\phi_l')\, \mathsf{e}^{\mathsf{j} \varphi_l},
\label{eq:dd_channel}
\end{equation}
where $\alpha_l$ is the complex gain of the $l$-th path  incorporating path-losses, small- and large-scale fading effects;
variables $\phi_l$   and    $\phi_l'$   are the azimuth  angle of arrival (AoA)
and angle of departure (AoD) for the $l$th path, respectively; and   $\mathbf{a}_{\rm T }(\cdot) \in \mathds{C}^{M_{\rm T}}$,
$\mathbf{a}_{\rm R }(\cdot)  \in \mathds{C}^{M_{\rm R}}$  represent the  transmit and receive array steering  vectors, respectively, which
depend on the antenna array geometry. Random phase $\varphi_l$ is associated with the delay of the $l$-th path.
Functions  $\mathsf{c}_{\rm T}(\cdot )$ and  $\mathsf{c}_{\rm R}(\cdot )$ represent the BS and UE antenna element directivity pattern,
respectively (all transmit antenna elements are assumed to have the same directivity pattern, and the same holds for the receive antenna elements).
Examples of transmit and receive antenna patterns are the uniform directivity pattern over a sector $[\phi_{\rm T}^{\rm l}, \phi_{\rm T}^{\rm u}]$, given by
% \begin{equation}
$
 \mathsf{c}_{\rm T}(\phi ) = 1$, when  $\phi \in [\phi_{\rm T}^{\rm l}, \phi_{\rm T}^{\rm u}]$
 and  $\mathsf{c}_{\rm T}(\phi ) = 0$, otherwise,
and likewise for $\mathsf{c}_{\rm R}(\phi )$.
Another baseline directivity pattern is advocated by 3GPP    \cite{3GPP_TS_37840_Rel12}
\begin{equation}
\mathsf{c}_{\rm T}(\phi) = 10^{\frac{\mathtt{G}_{\rm dB}}{20} + 
\max \left\{ -0.6 \left(\frac{\phi}{\phi_{\rm 3dB}}\right)^2, - \frac{\mathtt{A}_m}{20}\right\} },
\label{eq:ITU_antenna_pattern}
\end{equation}
with $\phi \in [-\pi,\pi)$, where $\mathtt{G}_{\rm dB}$ is  the maximum directional gain of the radiation element in dBi,
 $\mathtt{A}_m$ is the front-to-back ratio in dB, and
 $\phi_{\rm 3dB}$ is  the $3$dB-beamwidth.
 A common antenna array  architecture
is the uniform linear array (ULA) (w.r.t. $y$ axis) using only the azimuth angle; in this case
the BS steering vector (similarly for UE) is given by
\begin{equation}
 \mathbf{a}_{\rm T }(\phi)  = \sqrt\frac{1}{{M_{\rm T}}}\left[ 1~ \mathsf{e}^{-\mathsf{j}
 \frac{2 \pi d_{y}}{\lambda} \mathsf{sin}(\phi)}~\ldots~ \mathsf{e}^{-\mathsf{j}
 \frac{2 \pi d_y (M_{\rm T}-1)}{\lambda} \mathsf{sin}(\phi)}
 \right]^{\top},
 \label{eq:ULA_steering_vector}
\end{equation}
where $\lambda$ is the carrier wavelength, and
  $d_y$ is the distance between the antenna elements along the $y$ axis (usually $d_y = \lambda/2$).

The channel in~\eqref{eq:dd_channel} can be written more compactly  as
\begin{equation}
 \mathbf{H} = \mathbf{A}_{\rm R} \mathsf{diag}(\boldsymbol{\alpha})
  \mathbf{A}_{\rm T}^{\mathsf{H}},
%   = \left(\mathbf{A}_{\rm T}^*  \circ  \mathbf{A}_{\rm R} \right)
%   \boldsymbol{\alpha} .
  \label{eq:dd_channel_model_compact}
\end{equation}
 with matrices $
% \begin{subequations}\label{eq:compact_channel_matrix_elements}
% \begin{align}
\mathbf{A}_{\rm R} \triangleq  [ \mathsf{c}_{\rm R}(\phi_1 )  \mathbf{a}_{\rm R }(\phi_1)  ~ 
 \ldots~   \mathsf{c}_{\rm R}(\phi_L )\mathbf{a}_{\rm R }(\phi_L) ]$ and
%  \label{eq:AoA_matrix}\\
$\mathbf{A}_{\rm T} \triangleq   [\mathsf{c}_{\rm T}(\phi_1' ) \mathbf{a}_{\rm T }(\phi_1')  ~
 \ldots~ \mathsf{c}_{\rm T}(\phi_L' ) \mathbf{a}_{\rm T }(\phi_L') ]$ denoting  
   all transmit and receive steering vectors in compact form, respectively, while   vector
% \label{eq:AoD_matrix}\\
$\boldsymbol{\alpha} \triangleq \sqrt{\frac{M_{\rm T} M_{\rm R}}{L}}[\alpha_1 \mathsf{e}^{\mathsf{j} \varphi_1} ~  \ldots~
\alpha_L \mathsf{e}^{\mathsf{j} \varphi_L}]^{\top}$ collects the path-loss and phase shift coefficients.
% \end{align}
% \end{subequations}
Starting from the model in \eqref{eq:dd_channel_model_compact}, one can
come up with a sparse representation of the channel \cite{BaHaSaNo:10}.
First,  the angle space of AoA and AoD  is quantized by discretizing the
angular space. Let us denote these dictionaries $\mathcal{P}_{\rm T}$ and $\mathcal{P}_{\rm R}$ for AoDs  and AoAs, respectively.
Dictionary $\mathcal{P}_{\rm T}$ contains
$G_{\rm T}$ dictionary members, while $\mathcal{P}_{\rm R}$ contains $G_{\rm R}$
dictionary members.
One simple way of constructing these dictionaries is to use a uniform
grid of phases in an angular sector $\left[ a,  b \right) \subseteq [-\pi,\pi)$.
In that case, $\mathcal{P}_{\rm R} = \left\{ a  + \frac{ j \, (b-a)  }{G_{\rm R}+1} \right\}_{j=1}^{G_{\rm R}}$ and
$\mathcal{P}_{\rm T} = \left\{ a + \frac{  j \, (b-a) }{G_{\rm T}+1} \right \}_{j=1}^{G_{\rm T}}$.
For given dictionaries $\mathcal{P}_{\rm R}$ and $\mathcal{P}_{\rm T}$, dictionary  matrices
are defined
% \begin{subequations}\label{eq:dict_matrices}
% \begin{align}
%  \widetilde{\mathbf{A}}_{\rm R} \triangleq&~ \{ \mathsf{c}_{\rm R}(\phi) \, \mathbf{a}_{\rm R }(\phi) :
% \phi \in \mathcal{P}_{\rm R} \} \in \mathds{C}^{M_{\rm R} \times  {G}_{\rm R}}, \\
% \widetilde{\mathbf{A}}_{\rm  T} \triangleq&~ \{  \mathsf{c}_{\rm T}(\phi) \,\mathbf{a}_{\rm T }(\phi) :
%  \phi   \in \mathcal{P}_{\rm T} \}  \in \mathds{C}^{M_{\rm T} \times  {G}_{\rm T}},
% \end{align}
% \end{subequations}
\begin{align}
 \widetilde{\mathbf{A}}_{\rm R} \triangleq &~ \{ \mathsf{c}_{\rm R}(\phi) \, \mathbf{a}_{\rm R }(\phi) :
\phi \in \mathcal{P}_{\rm R} \} \in \mathds{C}^{M_{\rm R} \times  {G}_{\rm R}} ,  \\
\widetilde{\mathbf{A}}_{\rm  T} \triangleq &~  \{  \mathsf{c}_{\rm T}(\phi) \,\mathbf{a}_{\rm T }(\phi) :
 \phi    \in \mathcal{P}_{\rm T} \}  \in \mathds{C}^{M_{\rm T} \times  {G}_{\rm T}},
\end{align}
 which stand for an  overcomplete quantized approximation of the matrices $\mathbf{A}_{\rm R}$ and $\mathbf{A}_{\rm T}$,
 respectively. Hence, the channel matrix
 in the left-hand side of~\eqref{eq:dd_channel_model_compact} can be written, up to some quantization errors, as
\begin{equation}
 \mathbf{H} \approx  \widetilde{\mathbf{A}}_{\rm R}  \mathbf{G}
  \widetilde{\mathbf{A}}_{\rm T}^{\mathsf{H}},
  \label{eq:dd_channel_model_dictionary_compact1}
\end{equation}
where matrix  $  \mathbf{G}  \in   \mathds{C}^{ {G}_{\rm R} \times  {G}_{\rm T}}$
is an interaction matrix, whose $(j,k)$th element   is associated
with the $j$th and $k$th columns in $\widetilde{\mathbf{A}}_{\rm R}$
and $\widetilde{\mathbf{A}}_{\rm T}$, respectively -- if $[{\bf G}]_{j,k}\neq 0$, this
means that a propagation path associated with the $k$th angle in ${\cal P}_{\rm T}$
and the $j$th angle in ${\cal P}_{\rm R}$ is active.
In practice, the number of  active paths is typically very small
compared to the number of elements of $\mathbf{G}$ (i.e., $G_{\rm T}G_{\rm R}$).
Thus, the matrix $\mathbf{G}$ is in most cases very sparse  \cite{BaHaSaNo:10}.

 \begin{figure}%[!t]
\centering
\includegraphics[width=0.9\columnwidth]{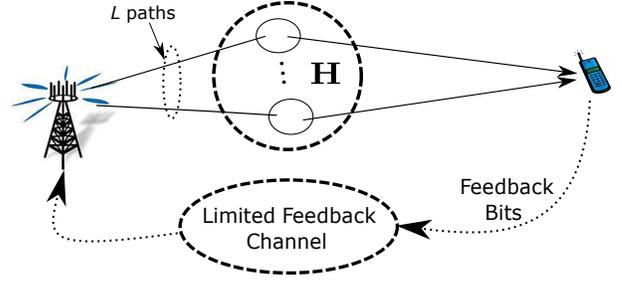}
\caption{ System model:  UE receives $\mathbf{y}$, employs one of the three limited feedback setups to compress the downlink channel matrix $\mathbf{H}$,   feeds back the  bits  to the BS through a limited feedback channel,
and the BS reconstructs  $\mathbf{H}$.}
\label{fig:figure_system_model}
\end{figure}

Stacking all columns   $\{\mathbf{y}_n\}_{n=1}^{N_{\rm tr}}$   in~\eqref{eq:baseband_equivalent} in a parallel fashion, we form matrix $\mathbf{Y} = [\mathbf{y}_1~ \mathbf{y}_2 ~\ldots~
\mathbf{y}_{N_{\rm tr}}]$.
Denoting $\mathbf{S} = [\mathbf{s}_1~\mathbf{s}_2~\ldots~\mathbf{s}_{N_{\rm tr}}]$
for the transmitted training symbol sequence
and $\mathbf{N} = [\mathbf{n}_1~\mathbf{n}_2~\ldots~\mathbf{n}_{N_{\rm tr}}]$ for the noise,  and
using  the channel matrix approximation in~\eqref{eq:dd_channel_model_dictionary_compact1},
the baseband signal in~\eqref{eq:baseband_equivalent}  can be written in a compact matrix form as
\begin{equation}
	\mathbf{Y} = \widetilde{\mathbf{A}}_{\rm R} \mathbf{G}
	\widetilde{\mathbf{A}}_{\rm T}^{\mathsf{H}} \mathbf{S}  + \mathbf{N}.
	\label{eq:baseband_equivalent_appr_compact}
\end{equation}
Applying the  vectorization    property $\mathsf{vec}(\mathbf{A B C}) = ( \mathbf{C}^{\top} \otimes \mathbf{A}) \mathsf{vec}(\mathbf{B})$
in Eq.~\eqref{eq:baseband_equivalent_appr_compact},
%of operator $\mathsf{vec}(\cdot)$,
%\begin{equation}
%	\mathsf{vec}(\mathbf{A B C}) = ( \mathbf{C}^{\top} \otimes \mathbf{A}) \mathsf{vec}(\mathbf{B}),
%\end{equation}
the baseband  received signal is given by
\begin{equation}
	\mathbf{y} =  \left((\mathbf{S}^{\top}   \widetilde{\mathbf{A}}_{\rm T}^{*})  \otimes
	\widetilde{\mathbf{A}}_{\rm R} \right) \mathbf{g} + \mathbf{n} = \mathbf{Q} \mathbf{g}
	+ \mathbf{n},
	\label{eq:baseband_equivalent_appr_vectorized}
\end{equation}
where $\mathbf{y} \triangleq \mathsf{vec}(\mathbf{Y})  \in \mathds{C}^{{M_{\rm R}  N_{\rm tr}}}$,
$\mathbf{g} \triangleq \mathsf{vec}(\mathbf{G})  \in \mathds{C}^{ G_{\rm T} G_{\rm R} }$,
$\mathbf{n} \triangleq \mathsf{vec}(\mathbf{N}) \in \mathds{C}^{{M_{\rm R}  N_{\rm tr}}}$,
and $\mathbf{Q} \triangleq (\mathbf{S}^{\top}   \widetilde{\mathbf{A}}_{\rm T}^{*})  \otimes
\widetilde{\mathbf{A}}_{\rm R}  \in \mathds{C}^{{M_{\rm R}  N_{\rm tr}} \times G_{\rm T} G_{\rm R}}$.
We define  $G \triangleq G_{\rm T} G_{\rm R} $   the joint (product) dictionary size.
This quantity plays a pivotal role on the performance of the algorithms considered, since it determines
the angle granularity of the dictionaries, which in turn determines the ultimate estimation error performance.
Fig.~\ref{fig:figure_system_model} provides a high-level overview of the system model. 
% and shows the three different limited feedback setups.
% {\color{blue} It is emphasized that the proposed setups  for downlink channel
% estimation with limited feedback provided in Sections~\ref{sec:UE_based_limited_feedback_sparse_channel_estimation}
% and~\ref{sec:Tx_based_limited_feedback_sparse_channel_estimation} can be applied
% to a multi-user setting, as the channel estimation process can be performed separately for each UE.}

\section{Angle Dictionary Construction Accounting for Antenna  Directivity Patterns}
\label{sec:angle_dict_construction}

Before introducing the proposed feedback schemes, let us consider the practical issue of quantizing the angular space.
Prior art on channel estimation employs the sparse representation in \eqref{eq:dd_channel_model_compact}
using uniformly discretized angles as dictionaries \cite{BaHaSaNo:10}.
However, a more appealing angle dictionary should take into consideration the
antenna directivity patterns, since the channel itself naturally reflects the directivity pattern.
% For instance, if a receive antenna has a spatial null, then it is impossible to receive a path coming
%  from the null direction, and likewise for transmit nulls.
  In this work we propose the following:  pack
more angles around the peaks of the antenna directivity pattern, because the dominant paths
will likely fall in those regions, and this is where we need higher angular resolution.
Denser discretization within high-antenna-power regions can reduce quantization errors more effectively compared to a
uniform quantization that ignores the directivity pattern.
% In this section, we propose
% a simple and easily implementable angle quantization technique that is based on the above rationale.
% As will be seen in the simulations, considering the directivity patterns substantially
% enhances the performance of channel estimation -- especially in the low SNR regime.

\begin{figure}[!t]
	\centering
	\includegraphics[width=0.8\columnwidth]{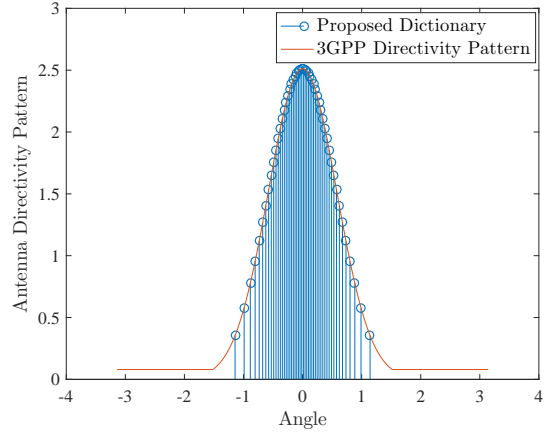} %\quad
	\caption{3GPP directivity pattern  along with function $\mathsf{q}$ applied on the proposed
		dictionary  using $a = -\pi$ and  $b = \pi$. The proposed angle dictionaries pack more points around higher values of $\mathsf{q}$.}
	\label{fig:3GPP_dipole_directivity_pattern_v1}
\end{figure}

To explain our approach, let  $\mathsf{q}  : [a,b) \longrightarrow \mathds{R}^{+}$ be a given
antenna directivity pattern function, which is assumed   continuous  over  $\phi \in [a,b)$
and suppose that we want to represent it using $N$ quantization points;
see Fig.~\ref{fig:3GPP_dipole_directivity_pattern_v1} for the 3GPP directivity pattern.
We define the cumulative function of $\mathsf{q}$, given by
$\mathsf{G}(\phi) \triangleq \int_{a}^{\phi} \mathsf{q}  (x) \mathsf{d}x $.
As the range  space of function $\mathsf{q} $  takes positive values,
its continuity implies that    $\mathsf{G}$ is monotone increasing.
Thus, the following set
\begin{equation}
\mathcal{C}_{\mathsf{q}} \triangleq  \left\{    \mathsf{G}(a) + \frac{n(\mathsf{G}(b)-  \mathsf{G}(a))}{N+1},
 \right\}_{n=1}^{N},
\label{eq:set_calC_g}
\end{equation}
partitions the range of  $\mathsf{G}$ in $N+1$ intervals of equal size.
By the definition of  $\mathsf{G}$, the set in~\eqref{eq:set_calC_g}
partitions  function $\mathsf{q} $ in $N+1$ equal area intervals.
Having the elements of set $ \mathcal{C}_{\mathsf{q}}$, we can find the
phases at which    $\mathsf{q}(\phi)$   is partitioned in
$N+1$ equal area intervals -- which means that we achieve our goal of putting denser grids
in the angular region where the ${\sf q}$ function has higher intensity.
These phases can be found as
\begin{equation}
\mathcal{F}_{\mathsf{q}} \triangleq
\left \{ \mathsf{G}^{-1}(y) \right \}_{y \in \mathcal{C}_{\mathsf{q}} },
\label{eq:set_calF_g}
\end{equation}
where $ \mathsf{G}^{-1}:[\mathsf{G}(a) , \mathsf{G}(b) ) \longrightarrow  [a,b)$ is the inverse (with respect to composition)
function of $\mathsf{G}$. Observe that $ \mathsf{G}^{-1}$ is a continuous, monotone increasing    function
since $\mathsf{G}$ is itself continuous and monotone increasing.
The discrete set $ \mathcal{F}_{\mathsf{q}}$ is a subset of $[a,b)$  and  concentrates more elements at points where
function $\mathsf{q}$ has larger values.
%Algorithm~\ref{alg:Dictionary_Construction} summarizes the proposed angle dictionary construction procedure.

Let us exemplify the procedure of constructing the angle dictionaries using the 3GPP antenna directivity pattern. 
%  Eq.~\eqref{eq:ITU_antenna_pattern} can be written equivalently as
% % \begin{equation}
% $
% \mathsf{q}(\phi)
% %= 10^{\frac{\mathtt{G}_{\rm dB}}{20}  +
% % \max \left\{ -0.6 \left(\frac{\phi}{\phi_{\rm 3dB}}\right)^2, - \frac{\mathtt{A}_m}{20}\right\} } \nonumber\\
%  =   10^{\frac{\mathtt{G}_{\rm dB}}{20} }  \,
% \mathsf{e}^{ \mathsf{ln}(10) \max \left\{ -0.6 \left(\frac{\phi}{\phi_{\rm 3dB}}\right)^2, - \frac{\mathtt{A}_m}{20}\right\} }. $
% \end{equation}
As the most general case  \cite{3GPP_TS_37840_Rel12}, we assume  $a \leq  - \phi_{\rm 3dB}\sqrt{\frac{\mathtt{A}_{\rm m}}{12}}$ and   $b \geq  \phi_{\rm 3dB}\sqrt{\frac{\mathtt{A}_{\rm m}}{12}}$.
The domain of $\mathsf{q}$ can be partitioned into 3 disjoint intervals as
%\begin{equation} $
$
[a, b) =   [a, -\phi_0) \cup [-\phi_0, \phi_0) \cup  [\phi_0, b)  $,
%\label{eq:3gpp_partition_interval}
%\end{equation}
with   $\phi_0 \triangleq  \phi_{\rm 3dB} \sqrt{\frac{\mathtt{A}_{\rm m}}{12}}$.
Using   $\mathsf{q}  (x) \equiv \mathsf{c}_{\rm T}(x)$  in Eq.~\eqref{eq:ITU_antenna_pattern}, applying the definition of cumulative function $\mathsf{G}(\phi) =  \int_{a}^{\phi} \mathsf{q}  (x) \mathsf{d}x$,
and using its continuity, we obtain   \cite{AlFuSidYaBl:17}
% \begin{figure*}[!b]
%\normalsize
%\hrulefill
% \setcounter{equation}{60}
\begin{equation}
\mathsf{G}(\phi)  \!=\!
\begin{cases}
  (\phi- a) 10^{ \frac{\mathtt{G}_{\rm dB}}{20}  - \frac{\mathtt{A}_m}{20} } ,& \!  \phi \in  [a, -\phi_0),\\
\mathsf{G}\!\left( - \phi_0 \right)  +   10^{\frac{\mathtt{G}_{\rm dB}}{20} }   \sqrt{\frac{\pi \, \phi_{\rm 3dB}^2 }{  \mathsf{ln}(10) \, 2.4} } \cdot  & \\ 
\left(  \mathsf{erf} \!\left( \sqrt{ \frac{\mathsf{ln}(10) \, 0.6\, \mathtt{A}_m}{12}} \right) + \right. & \\
\left.  \mathsf{sign}(\phi)    \mathsf{erf} \!\left( \sqrt{ \frac{\mathsf{ln}(10) \, 0.6 }{\phi_{\rm 3dB}^2}}  |\phi| \right) \right) ,
& \!   \phi \in [-\phi_0, \phi_0) , \\
\mathsf{G}\!\left(  \phi_0 \right)  +      \left(\phi - \phi_0 \right) 10^{ \frac{\mathtt{G}_{\rm dB}}{20}  - \frac{\mathtt{A}_m}{20} }   , & \! \phi \in  [\phi_0, b),
\end{cases}
\label{eq:cumulative_function_3gpp}
\end{equation}
where $\mathsf{erf}(x)   \frac{\sqrt{\pi}}{2} = \int_{0}^{x} \mathsf{e}^{-t^2} \mathsf{d}t$ was utilized.
Upon defining  $y^{-} \triangleq \mathsf{G}(-\phi_0)$, $y_0 \triangleq  \mathsf{G}(0) $,
and   $y^{+} \triangleq \mathsf{G}(\phi_0)$, the inverse of  $\mathsf{G}(\cdot)$ can be calculated
using Eq.~\eqref{eq:cumulative_function_3gpp} in closed form as
\begin{equation}
\mathsf{G}^{-1}\!(y)\!  = \!
\begin{cases}
y \, 10^{\frac{\mathtt{A}_m}{20} - \frac{\mathtt{G}_{\rm dB}}{20}  }    + a ,&  \! y \in  [0, y^{-}), \\
-\frac{ \mathsf{erf}^{-1}\!\left(   \frac{2 \sqrt{\mathsf{ln}(10) \, 0.6} }{\phi_{\rm 3dB} \sqrt{\pi}}
	 ( y_0 - y ) 10^{- \frac{\mathtt{G}_{\rm dB}}{20}  }   \right)  }{\frac{\sqrt{\mathsf{ln}(10) \, 0.6}}{\phi_{\rm 3dB}} }  ,&
\! y \in  [y^{-},y_0), \\
\frac{ \mathsf{erf}^{-1}\!\left(   \frac{2 \sqrt{\mathsf{ln}(10) \, 0.6} }{\phi_{\rm 3dB} \sqrt{\pi}}
	  (   y - y_0  ) 10^{- \frac{\mathtt{G}_{\rm dB}}{20}  }  \right) }{\frac{\sqrt{\mathsf{ln}(10) \, 0.6}}{\phi_{\rm 3dB}} }  ,&
\! y \in  [ y_0, y^{+}), \\
\phi_0 +  (y -y^{+})    10^{\frac{\mathtt{A}_m}{20} - \frac{\mathtt{G}_{\rm dB}}{20}  }  ,&
\! y \in  [  y^{+}, \mathsf{G}(b)),
\end{cases}
\label{eq:cumulative_function_3gpp_inv}
\end{equation}
where $\mathsf{erf}^{-1}(\cdot)$ is the inverse (with respect to composition) function
of $\mathsf{erf}(\cdot)$, and is well tabulated by several software packages, such as Matlab.
The definition of inverse function in~\eqref{eq:cumulative_function_3gpp_inv}
for interval $  [a,b)$, such that $[-\phi_0, \phi_0) \subseteq [a,b) \subseteq [-\pi , \pi) $,
is   the most general case.  As one can see in Fig.~\ref{fig:3GPP_dipole_directivity_pattern_v1},
the point density of this quantization of the angular space indeed reflects the selectivity of the antenna directivity pattern, as desired.

\section{UE-based Baseline Limited Feedback Sparse Channel Estimation}
\label{sec:UE_based_limited_feedback_sparse_channel_estimation}

This section presents a baseline limited feedback setup where UE
estimates the sparse channel and sends back the support along with the
coarsely quantized nonzero elements of the estimated sparse channel $\mathbf{g}$.

\subsection{Channel  Estimation and Support Identification at UE}
\label{subsub:sparse_channel_estim_supp_ident}

The inherent sparsity of $\mathbf{g}$ in~\eqref{eq:baseband_equivalent_appr_vectorized} suggests the following formulation to
recover it  at  UE
\begin{align}
  \min_{\mathbf{g} \in \mathds{C}^{G} :  \|\mathbf{g} \|_0  \leq L }
 \!\left\{\frac{1}{2}  \|\mathbf{y} - \mathbf{Q} \mathbf{g} \|_2^2   \right\}.
  \label{eq:optim_LS_l0_reg}
% {\rm subject~ to} & ~~~~ \|\mathbf{g} \|_0  \leq L .
% \end{split}
\end{align}
% \begin{align}
%   \label{eq:optim_LS_l0_reg}
% \begin{split}
%      \underset{\mathbf{g}  \in \mathds{C}^{ G  }} {\rm minimize}&~~~~
%  \frac{1}{2}  \|\mathbf{y} - \mathbf{Q} \mathbf{g} \|_2^2  \\
% {\rm subject~ to} & ~~~~ \|\mathbf{g} \|_0  \leq L .
% \end{split}
% \end{align}
The optimization problem in~\eqref{eq:optim_LS_l0_reg} is a non-convex combinatorial problem.
Prior art  in  compressed sensing (CS) optimization literature has attempted
to solve~\eqref{eq:optim_LS_l0_reg} using approximation algorithms,
such as  orthogonal matching pursuit (OMP)
\cite{TroGil:07}, iterative hard thresholding (IHT) \cite{Blum:12}, and
many others; see \cite{ChDiRaKi:15} and  references therein.
OMP-based algorithms are preferable for sparse channel estimation, due to their favorable performance-complexity trade-off
\cite{ChDiRaKi:15}. OMP admits simple and even real-time implementation,
and its run-time complexity can be further reduced by caching the QR factorization of matrix
$\mathbf{Q}$ \cite{TroGil:07}.  For completeness, the pseudo code for OMP is provided in Algorithm~\ref{alg:OMP}.
 For a detailed discussion regarding the implementation details and  performance guarantees of the OMP algorithm the reader is 
referred to \cite{FoucRau:13}.

{ \footnotesize
\begin{algorithm}[!t]
\caption{\small Channel Estimation and Support Identification at UE}
{\small \textbf{Input:} $  \mathbf{Q},  \mathbf{y}$, $\overline{L}$  }
\begin{algorithmic} [1]
 \State{\small $t=0:$ Initialize $\mathbf{r} =  \mathbf{y}$, $\mathcal{S}_{\widehat{\mathbf{g}}} = \emptyset$}
\While {\small  $\| \mathbf{Q}^{\mathsf{H}} \mathbf{r}\|_{\infty} > \varepsilon$ {\bf and} $t <  \overline{L}$ }
\State{\small $t:=t+1$}
\State{\small $ \mathbf{p}   =\mathbf{Q}^{\mathsf{H}} \mathbf{r} $ }
\State{\small $n^{\star}   =  \arg \max_{n = 1,2,\ldots G} \{| p_n| \}$  }
\State{\small $ \mathcal{S}_{\widehat{\mathbf{g}}} : = \mathcal{S}_{\widehat{\mathbf{g}}} \cup  n^{\star}$}
\State{\small $\widehat{\mathbf{g}}_{\mathcal{S}_{\widehat{\mathbf{g}}}^{\mathtt{C}}}  = \mathbf{0}$ and $
\widehat{\mathbf{g}}_{\mathcal{S}_{\widehat{\mathbf{g}}}}  = \mathbf{Q}_{:,\mathcal{S}_{\widehat{\mathbf{g}}}}^{\dagger} \mathbf{y}$}
\State{\small $  \mathbf{r}  = \mathbf{y} - \mathbf{Q}\, \widehat{\mathbf{g}} $ }
  \EndWhile
\end{algorithmic}
{\small \textbf{Output:}  $\widehat{\mathbf{g}}$, $\mathcal{S}_{\widehat{\mathbf{g}}}$}
\label{alg:OMP}
\end{algorithm}
}

\subsection{Scalar Quantization and Limited Feedback}
\label{subsubsec:scalarQuant}

After estimating the sparse vector $\widehat{\mathbf{g}}$  associated with an estimate of interaction matrix  $\widehat{\mathbf{G}}$ a  simple
feedback technique is to  send coarsely quantized non-zero elements of  $\widehat{\mathbf{g}}$, along with the corresponding indices.
In this work we make use of Lloyd's scalar quantizer to quantize the non-zero elements
of  $\widehat{\mathbf{g}}$, and we denote the scalar quantization operation $\mathsf{SQ}(\widehat{\mathbf{g}})$.
Upon receiving the bits associated with the non-zero indices and elements    of  $\widehat{\mathbf{g}}$,
i.e., $\mathcal{S}_{\widehat{\mathbf{g}}}$ and $\mathsf{SQ}(\widehat{\mathbf{g}})$, the BS reconstructs channel matrix $\widehat{\mathbf{H}}$
via~\eqref{eq:dd_channel_model_dictionary_compact1},   provided it has 
perfect knowledge of SQ threshold values. As the channel model in~\eqref{eq:baseband_equivalent_appr_vectorized} has  sparse
structure comprising $L$ non-zero elements, for suitably designed $\mathbf{Q}$ and a sufficient number of training symbols, this approach
tends to yield a channel estimate comprising $\mathcal{O}(L)$ non-zero elements.

Using a $Q$-bit real scalar quantizer, each non-zero element of complex vector $\widehat{\mathbf{g}}$ can be represented
using $\lceil\mathsf{log}_2 G\rceil +  2Q$ bits, where the first term accounts for index coding, and the second for coding the real and imaginary parts.
Hence,
the total number of feedback bits to estimate the interaction matrix $\mathbf{G}$
at the BS,  scales with
$\mathcal{O}(L ( \lceil\mathsf{log}_2 G\rceil   + 2Q))$. In the worst case, OMP iterates
$\overline{L}$ times, offering worst case feedback overhead $ \overline{L} ( \lceil\mathsf{log}_2 G\rceil   + 2Q) $.
Note that the  number of feedback bits of the proposed  UE-based baseline limited  feedback algorithm
is independent of $M_{\rm T}$.

% Thus, for small number of paths $L$, and thus small upper bound $\overline{L}$,
% the
% can be significantly less than the number of typical state-of-the-art VQ feedback schemes, which
% require that the number of feedback bits scales up linearly with $M_{\rm T}$.

\section{BS-based Limited Feedback Sparse Channel Estimation}
\label{sec:Tx_based_limited_feedback_sparse_channel_estimation}

In order to reduce the feedback overhead without irrevocably sacrificing our ability to recover accurate CSI at the BS,
we propose to apply a {\em pseudo-random} dimensionality-reducing linear operator $\mathbf{P}^{\mathsf{H}}$ to $\mathbf{y}$.  The outcome is  quantized with a very simple sign
quantizer, whose output is fed  back to the BS through a low-rate channel. More precisely, the BS receives
\begin{equation}
   \mathbf{b}_{\Re} + \mathsf{j} \,\mathbf{b}_{\Im} =
\mathsf{sign}(\Re\!\left(\mathbf{P}^{\mathsf{H}} \mathbf{y}\right))  +
\mathsf{j}  \, \mathsf{sign}\!\left(\Im\!\left(\mathbf{P}^{\mathsf{H}} \mathbf{y}\right)\right),
\label{eq:received_feed_Rx_to_Tx}
\end{equation}
where $\mathbf{P} \in \mathds{C}^{M_{\rm R}N_{\rm tr} \times N_{\rm fb}}$,
with  $N_{\rm fb} \leq M_{\rm R}N_{\rm tr}$.

To facilitate operating in the more convenient real domain, consider the following definitions
\begin{subequations}\label{eq:real_to_complex}
\begin{align}
 \mathbf{C}_{\Re}^{\top}  &\triangleq \left[ {\Re}(\mathbf{Q}^{\mathsf{H}} \mathbf{P})^{\top}  ~  {\Im}(\mathbf{Q}^{\mathsf{H}} \mathbf{P})^{\top} \right],
   \label{eq:matrix_C_real}\\~~
  \mathbf{C}_{\Im}^{\top} & \triangleq \left[- {\Im}(\mathbf{Q}^{\mathsf{H}} \mathbf{P})^{\top}  ~ {\Re}(\mathbf{Q}^{\mathsf{H}} \mathbf{P})^{\top}
 \right]   , \label{eq:matrix_C_imag}\\
 {\mathbf{C}} &\triangleq [ \mathbf{C}_{\Re}~~ \mathbf{C}_{\Im}] = \left[\mathbf{c}_1~ \mathbf{c}_2
\ldots ~\mathbf{c}_{2N_{\rm fb}} \right] \in   \mathds{R}^{2G \times 2 N_{\rm fb}} ,   \label{eq:matrix_C}\\
\mathbf{x}^{\top} &\triangleq \left[ {\Re}(\mathbf{g})^{\top}  ~  {\Im}(\mathbf{g})^{\top} \right] \in  \mathds{R}^{2G}, ~~    \label{eq:vector_x}\\
\mathbf{b}^{\top} &\triangleq \left[  \mathbf{b}_{\Re}^{\top}
~\mathbf{b}_{\Im} ^{\top} \right]^{\top}  = [b_1~b_2~\ldots~b_{2N_{\rm fb}}] \in  \mathds{R}^{2 N_{\rm fb}}   \label{eq:vector_b} ,
  \\
\mathbf{z}^{\top}  &  \triangleq  \left[  \mathbf{z}_{\Re}^{\top}
~\mathbf{z}_{\Im} ^{\top} \right]^{\top}=  [z_1~z_2~\ldots~z_{2N_{\rm fb}}]  \in  \mathds{R}^{2 N_{\rm fb}}  \label{eq:vector_z},
\end{align}
\end{subequations}
with $ \Re\!\left( \mathbf{P}^{\mathsf{H}} \mathbf{Q} \mathbf{g}\right) =  \mathbf{C}_{\Re}^{\top} \mathbf{x}$,
$  \Im\!\left( \mathbf{P}^{\mathsf{H}} \mathbf{Q} \mathbf{g}\right) =  \mathbf{C}_{\Im}^{\top} \mathbf{x}$,
$ \mathbf{z}_{\Re} =  \Re \!\left(\mathbf{P}^{\mathsf{H}} \mathbf{n}\right)$, and  $ \mathbf{z}_{\Im} =
 \Im \! \left(\mathbf{P}^{\mathsf{H}} \mathbf{n}\right)$.
Using the above,  along with~\eqref{eq:received_feed_Rx_to_Tx},
the received feedback bits at the BS are given by
\begin{align}
% \mathbf{b} &= \mathsf{sign}\!\left(\mathbf{C}^{\top} \mathbf{x} + \mathbf{z}\right)
\label{eq:real_equivalent1}
    b_i  = \mathsf{sign}\!\left(\mathbf{c}_i^{\top} \mathbf{x} +  z_i\right),~~i=1,2,\ldots, 2N_{\rm fb}.
\end{align}
The objective at the BS is to estimate $\mathbf{x}$, given $\mathbf{b}$
and $\mathbf{C}$. If the complex vector $\mathbf{g}$ has $L$ non-zero elements, then the
real vector $\mathbf{x}$ has up to $2L$ non-zero elements. More precisely, vector $\mathbf{x}$ has
$L$ active (real, imaginary) element pairs, i.e., it exhibits group-sparsity of order $L$, where the groups are predefined pairs here.
In our experiments, we have noticed that the distinction hardly makes a difference in practice. In the sequel, we therefore
drop group sparsity in favor of simple $2L$ sparsity.

 It should be  noted that the number of feedback bits $N_{\rm fb}$ is controlled by the dimension
 of $\mathbf{P}$, which is determined by the designer to balance channel estimation accuracy versus the feedback rate.
  As  $\|\mathbf{x}\|_0 \leq 2L$, from   compressive sensing theory  we know that
the number of measurements to recover $\mathbf{x}$ is lower bounded by $4L$ \cite{FoucRau:13}. In practice, depending on  the examined cellular setting,
it is usually easy to have a rough idea of $L$ \cite{KamKhAltDebKam:14}.

\subsection{Single-Bit Compressed   Sensing  Formulation}
\label{subsec:CS_formulation}

Single-bit compressed sensing (CS) has attracted significant attention in the compressed sensing literature
 \cite{BouBar:08,   PlanVer:13_a,  JacLasBouBar:13, ZhaYiJin:14}, where the goal is to reconstruct a sparse signal
 from single-bit measurements. Existing single-bit CS algorithms make the explicit assumption that $\|\mathbf{x}\|_2=1$
 \cite{BouBar:08}, or  $\|\mathbf{x}\|_2 \leq 1$ \cite{PlanVer:13_a,ZhaYiJin:14}.
Thus,  the solution of  single-bit CS problems is always a sparse vector on  a unit hypersphere.
In our context,  we seek a sparse $\mathbf{x}$ that yields maximal agreement between the observed and the reconstructed signs.
This suggests the following formulation
\begin{equation}\label{eq:true_obj}
\widehat{\mathbf{x}} = \arg \min_{\mathbf{x}  \in \mathds{R}^{ 2G  }}  \left \{
-\sum_{i=1}^{2N_{\rm fb}} \mathsf{sign}({\bf c}_i^{\top}{\bf x}) \,b_i+\zeta\|{\bf x}\|_0 \right\},
\end{equation}
where $\zeta>0$ is a regularization parameter  that controls the sparsity of
the optimal solution. Unfortunately the optimization problem in~\eqref{eq:true_obj}
is non-convex and requires exponential complexity to be solved to global optimality.
In addition, notice that the scaling of $\mathbf{x}$ cannot be determined from \eqref{eq:true_obj}: if $\mathbf{x}$ is an optimal solution, so is $c \mathbf{x}$ for any $c > 0$.
Therefore, the following convex surrogate of problem~\eqref{eq:true_obj} is considered
\begin{equation}
\widehat{\mathbf{x}}=\arg \min_{\mathbf{x}  \in \mathds{R}^{ 2G  }: \|\mathbf{x}\|_2 \leq  R_2}  \left\{ - \mathbf{x}^{\top} \mathbf{C}\, \mathbf{b} + \zeta\, \| \mathbf{x} \|_1 \right\},
\label{eq:optimization_binary_meas_CS_1}
\end{equation}
where $R_2$ is an upper bound on the norm of $\mathbf{x}$, which also prevents meaningless scaling up of ${\bf x}$
when $\zeta$ is small.  We found that setting $R_2$ to be on the same order of magnitude with
$\mathtt{P}_{\alpha} = \sqrt{\sum_{l=1}^L v_l}$ works very well,  where  $v_l = \mathbb{E}[|\alpha_l|^2]$; note that quantity $\mathtt{P}_{\alpha}$
expresses the aggregated  power of the wireless channel gain coefficients in Eq.~\eqref{eq:dd_channel}.
The cost function in \eqref{eq:optimization_binary_meas_CS_1} is known to be an effective surrogate of the one in \eqref{eq:true_obj},
both in theory and in practice. If the  elements of $\mathbf{C}$ are drawn from a Gaussian distribution, the formulation 
in~\eqref{eq:optimization_binary_meas_CS_1} will recover $2L$-sparse $\mathbf{x}$ on the unit hypersphere (i.e., $R_2=1$)
with $\epsilon$-accuracy using $\mathcal{O}(\frac{2 L \mathsf{log} G}{\epsilon^4})$ measurements \cite{ ZhaYiJin:14}.

Interestingly,  %unlike its real-valued input counterpart,
problem \eqref{eq:optimization_binary_meas_CS_1} admits closed-form solution, given by \cite{ZhaYiJin:14}
\begin{equation}
\widehat{\mathbf{x}} =
\begin{cases}   \mathbf{0}, &  \|    \mathbf{C}\, \mathbf{b} \|_{\infty} \leq \zeta,
\\
\frac{R_2\,  \mathsf{T} (   \zeta  ;  \,    \mathbf{C}\, \mathbf{b})}{\left\| \mathsf{T}(   \zeta  ;  \,     \mathbf{C}\, \mathbf{b}  ) \right\|_2}, & {\rm otherwise},
\end{cases}
\label{eq:CS_estimate_x}
\end{equation}
where for $v>0$,   $\mathsf{T}(v; \cdot) : \mathds{R}^{2G} \longrightarrow \mathds{R}^{2G}  $ denotes the shrinkage-thresholding operator,
given by
\begin{equation}
 [\mathsf{T}( v; \mathbf{x} ) ]_i= (|x_i| -v)_{+} \, \mathsf{sign}(x_i), ~~i=1,2,\ldots, 2G.
\end{equation}

The overall computational cost of computing~\eqref{eq:CS_estimate_x}  is $\mathcal{O}( N_{\rm fb} G)$.
A key advantage of the adopted CS method  is that it is a closed-form expression, and thus it is very easily
implementable in real-time.

\subsection{Sparse Maximum-Likelihood Formulation}
\label{subsec:sparse_ML_formulation}

Let $\mathbf{P}$ be a semi-unitary matrix, i.e., $\mathbf{P}^{\mathsf{H}} \mathbf{P} = \mathbf{I}_{N_{\rm fb}}$.
Because vector $\mathbf{n}$ is a circularly-symmetric complex Gaussian vector, the statistics of the noise vector
$\mathbf{z}$  are $\mathcal{N}\left(\mathbf{0}, \sigma_z^2 \, \mathbf{I}_{2N_{\rm fb}}\right)$, where $\sigma_z^2 = \frac{\sigma^2}{2}$.
So, each $b_i = \mathsf{sign}(\mathbf{c}_i^{\top} \mathbf{x}+ z_i)$ is a Rademacher random variable (RV)
with parameter ${\rm  Pr}(b_i=1) =1 - {\rm  Pr}(b_i=-1) =  {\rm  Pr}(\mathbf{c}_i^{\top} \mathbf{x}+ z_i > 0) =
\mathsf{Q}\!\left(- \frac{\mathbf{c}_i^{\top} \mathbf{x}}{\sigma_z}\right)$.
In addition to that, due to the fact that $\mathbf{z}$'s covariance matrix is diagonal,  all $\{b_i\}_{i=1}^{2N_{\rm fb}}$
are independent of each other.

In the proposed sparse maximum-likelihood (ML) formulation, the sparse  channel parameter vector is estimated
 by maximizing the regularized  log-likelihood of the (sign) observations,
 $\mathbf{b}$, given $\mathbf{x}$. Using the independence of $\{b_i\}_{i=1}^{2 N_{\rm fb}}$,
the sparse ML problem can be formulated as \cite{TsaJaSidOtt:13}
\begin{align}
  \label{eq:optim_ML_l1_reg}
\underset{ \mathbf{x}  \in \mathds{R}^{ 2G  }} {\rm inf}   \left\{  -
 \sum_{i=1}^{2  N_{\rm fb}}  \mathsf{ln}\;\mathsf{Q} \! \left( -b_{i} \frac{\mathbf{c}_{i}^{\top}\mathbf{x}}{\sigma_z} \right)
    + \zeta \| \mathbf{x}\|_1 \right\},
\end{align}
where $\zeta \geq 0$ is a  tuning  regularization parameter that controls the sparsity
of the solution. 
 Let us denote   $\mathsf{f}(\mathbf{x}) \triangleq - \sum_{i=1}^{2  N_{\rm fb}}
\mathsf{ln}\;\mathsf{Q} \! \left( -b_{i} \frac{\mathbf{c}_{i}^{\top}\mathbf{x}}{\sigma_z} \right)$
and $\mathsf{h}(\mathbf{x}) \triangleq \mathsf{f}(\mathbf{x}) + \zeta \| \mathbf{x}\|_1$.
The above  is a convex optimization problem  since the $\mathsf{Q}$-function is log-concave
\cite[p.~104]{BoydVand:04}. According to the Weierstrass theorem, the minimum   in~\eqref{eq:optim_ML_l1_reg}
always exists since the objective, $\mathsf{h}(\cdot)$, is a coercive function,
meaning that for any sequence $\left\{\mathbf{x}^{(t)}\right\}_{t=1}^{\infty}$, such that $\|\mathbf{x}^{(t)}\|
\longrightarrow \infty$,  $\lim_{t \rightarrow \infty}\mathsf{h} (\mathbf{x}^{(t)}) = \infty$  holds true \cite[p.~495]{Bert:15}.
A choice for $\zeta $ that  guarantees that  the all-zero vector is not solution of~\eqref{eq:optim_ML_l1_reg}
 is $\zeta \leq \| \nabla  \mathsf{f}(\mathbf{0})  \|_{\infty}$
 (the proof of this claim relies on a  simple application of optimality conditions using subdifferential
 calculus \cite{Bert:15}), where the gradient of $ \mathsf{f}(\cdot)$ is given by \cite{MehSid:14_a}
\begin{equation}
\nabla  \mathsf{f}(\mathbf{x})  =  - \sum_{i = 1}^{2 N_{\rm fb}}
\frac{  b_{i}\, \mathsf{e}^{-\frac{\left(\mathbf{c}_{i}^{\top}\mathbf{x} \right)^2}{2\sigma_z^2}}}
{\sqrt{2 \pi } \sigma_z   \mathsf{Q}\!\left( -b_{i} \frac{\mathbf{c}_{i}^{\top}\mathbf{x}}{\sigma_z} \right) } \mathbf{c}_{i}
   \label{eq:l1_reg_gradient_f_ML}  .
\end{equation}
%   This gives an upper bound on $\zeta$.
%  In practice parameter $\zeta$  can be tuned from historical data.

% \begin{remark} \normalfont
It is worth noting that the minimizer of problem~\eqref{eq:optim_ML_l1_reg} can be also viewed
 as  the maximum a-posteriori probability (MAP) estimate of $\mathbf{x}$ under the assumption
 that the elements of vector $\mathbf{x}$ are independent of each other and
 follow a Laplacian distribution.
% \end{remark}

%  To solve the above problem the optimal first-order method of Nesterov will be adopted \cite{Nes:04}.
The Hessian of $ \mathsf{f}(\cdot)$ is given by \cite{MehSid:14_a}
\begin{equation}
\nabla^2  \mathsf{f}(\mathbf{x})  =    ~ \mathbf{C}\, \mathsf{diag}  (\mathbf{m}(\mathbf{x}))\, \mathbf{C}^{\top}    , \label{eq:gradient_f_ML}
\end{equation}
where the elements of vector $ \mathbf{m}(\cdot) $ are  given by
\begin{equation}
  {m}_{i}(\mathbf{x}) \! =  \!\frac{  \mathsf{e}^{-\frac{\left(\mathbf{c}_{i}^{\top}\mathbf{x} \right)^2}{\sigma_z^2}}}
{  2   \pi  \sigma_z^2 \left[\mathsf{Q}\!\left( - \frac{ b_{i} \, \mathbf{c}_{i}^{\top}\mathbf{x}}{\sigma_z} \right)  \right]^2}
\!+\!
 \frac{    b_{i} \,( \mathbf{c}_{i}^{\top}\mathbf{x}) \, \mathsf{e}^{-\frac{\left(\mathbf{c}_{i}^{\top}\mathbf{x} \right)^2}{2\sigma_z^2}}}
{ \sqrt{ 2  \pi} \sigma_z^3 \, \mathsf{Q}\!\left( - \frac{ b_{i} \, \mathbf{c}_{i}^{\top}\mathbf{x}}{\sigma_z} \right)   }
\label{eq:diag_Hess_f_ML_m_i},
\end{equation}
$i = 1,2,\ldots, 2N_{\rm fb}$.
Having calculated the Hessian,   due to Cauchy-Swartz inequality for matrix norms
\begin{align}
\| \nabla^2  \mathsf{f}(\mathbf{x}) \|_2 \leq &~  \|  \mathbf{C}\|_2
\| \mathsf{diag}  (\mathbf{m}(\mathbf{x})) \|_2 \|  \mathbf{C}^{\top}\|_2 \nonumber \\
=&~  \|  \mathbf{C} \|_2^2    \|\mathbf{m}(\mathbf{x})\|_{\infty}
\triangleq  \mathsf{L}(\mathbf{x}) ,~\forall \mathbf{x} \in \mathds{R}^{2G} . \label{eq:l1_reg_bound2_hessian_f}
\end{align}
It is noted that   for bounded $\|\mathbf{x}\|_1$,   $\mathsf{L}(\mathbf{x})$ is also bounded.

An accelerated gradient method for the  $l_1$-regularized problem in~\eqref{eq:optim_ML_l1_reg} is utilized, where
 sequences $\{ \mathbf{x}^{(t)}\}$ and $\{ \mathbf{u}^{(t)}\}$  are generated  according to    \cite{BeckTeb:09}
 \begin{subequations}\label{eq:FISTA_equation}
\begin{align}
\mathbf{x}^{(t+1)}& = \mathsf{T}\!\left(  \frac{\zeta}{  \mathsf{L}\!\left(\mathbf{u}^{(t)}\right)} ;  \;
\mathbf{u}^{(t )} -
\frac{1}{  \mathsf{L}\!\left(\mathbf{u}^{(t)}\right) } \nabla  \mathsf{f}\!\left(\mathbf{u}^{(t)}\right) \right),
\label{eq:ML_update_x_t} \\
\beta^{(t+1)} & = \frac{1 + \sqrt{1 + 4 \left(\beta^{(t)} \right)^2}}{2}  \label{eq:ML_update_beta_t},\\
\mathbf{u}^{(t+1)} &= \mathbf{x}^{(t+1)}+ \frac{\beta^{(t)} - 1}{\beta^{(t+1)} }  \left(\mathbf{x}^{(t+1)} -  \mathbf{x}^{(t)}\right) .
\label{eq:ML_update_u_t}
\end{align}
\end{subequations}
For bounded   $\mathsf{L}(\cdot)$, which holds in our case, the sequence generated by
 updates in~\eqref{eq:FISTA_equation}  
converges to an $\epsilon$-optimal solution (a neighborhood of the optimal solution with diameter $\epsilon$) 
using at most $\mathcal{O}(1/\sqrt{\epsilon})$ iterations \cite{BeckTeb:09}.

 {
\small
\begin{algorithm}[!t]
\caption{\small Limited Feedback Sparse ML Channel Estimation}
{\small \textbf{Input:} $  \mathbf{C},  \mathbf{b}$, $\zeta$} %$\mathtt{tol}$, $\mathtt{I}_{\rm max}$
\begin{algorithmic} [1]
\State{\small Precompute $ \| \mathbf{C}\|_2^2$  }
\State{\small $t=0:$ Initialize $\beta^{(0)}  = 1$, $\mathbf{u}^{(0)} = \mathbf{x}^{(0)} \in \mathds{R}^{2G}$}
\While { \small Stopping criterion is not reached }
\State{\small $\mathsf{L}^{(t)}  =  \| \mathbf{C}\|_2^2   \|\mathbf{m}( \mathbf{u}^{(t)} )\|_{\infty}$}
\State{\small $ \mathbf{x}^{(t+1)}   = \mathsf{T}\!\left(  \frac{\zeta}{  \mathsf{L}^{(t)}} ;  \;
\mathbf{u}^{(t )} -
\frac{1}{\mathsf{L}^{(t)}} \nabla  \mathsf{f}\!\left(\mathbf{u}^{(t)}\right) \right)  $ }
\State{\small $\beta^{(t+1)}   = \frac{1 + \sqrt{1 + 4 \left(\beta^{(t)} \right)^2}}{2}$  }
\State{\small$ \mathbf{u}^{(t+1)} = \mathbf{x}^{(t+1)}+ \frac{\beta^{(t)} - 1}{\beta^{(t+1)} }  \left(\mathbf{x}^{(t+1)} -  \mathbf{x}^{(t)}\right)$}
\If {\small $  \nabla \mathsf{f}\!\left(\mathbf{u}^{(t)} \right)^{\top}
\left(\mathbf{x}^{(t+1)} - \mathbf{x}^{(t)}\right)  > 0  $ }
\State{\small $\beta^{(t+1)}  = 1$,   $  \mathbf{u}^{(t+1)}  = \mathbf{x}^{(t+1)} $}
  \EndIf
\State{\small $t:=t+1$}
  \EndWhile
\end{algorithmic}
{\small \textbf{Output:}  $\widehat{\mathbf{x}} = \mathbf{x}^{(t)}  $}
\label{alg:Limited_Feedback_Sparse_ML}
\end{algorithm}
}

Algorithm~\ref{alg:Limited_Feedback_Sparse_ML} illustrates the proposed  first-order $l_1$-regularization algorithm
incorporating Nesterov's   extrapolation method. In addition, an adaptive restart mechanism \cite{ODonCan:15}
is  utilized in order to further speed up the convergence rate.
Experimental  evidence on our problems shows that it  works remarkably well.
At line (1),  quantity $\|\mathbf{C}\|_2^2$ is precomputed, requiring  $\mathcal{O}( N_{\rm fb}^2 \, G)$ 
arithmetic operations.
The per iteration complexity of the proposed algorithm is $\mathcal{O}( N_{\rm fb} \, G)$ due to the evaluation of
$   \nabla  \mathsf{f}(\mathbf{u}^{(t)}) $ and $\mathbf{m}(\mathbf{u}^{(t)})$  
at lines 4 and 5, respectively. In the worst case, MLE-reg algorithm iterates
$I_{\rm max}$ times offering total computational cost $\mathcal{O}((I_{\rm max} + N_{\rm fb}) N_{\rm fb} \, G)$.  
Note that such complexity is linear in  $G$, and thus,  affordable at a typical BS.

To reconstruct an estimate of the downlink channel,
the BS obtains $\widehat{\mathbf{g}}$ from $\widehat{\mathbf{x}}$
as $\widehat{\mathbf{g}} = \widehat{\mathbf{x}}_{1:G} + \mathsf{j} \widehat{\mathbf{x}}_{G+1:2G} $
and   forms an estimate of the interaction  matrix $\widehat{\mathbf{G}}$ using the inverse of the vectorization operation,
i.e., $\widehat{\mathbf{G}} = \mathsf{unvec}(\widehat{\mathbf{g}})$.
With   $\widehat{\mathbf{G}}$ available, the downlink channel matrix can be estimated 
 as  $\widehat{\mathbf{H}} =
 \widetilde{\mathbf{A}}_{\rm R}  \widehat{\mathbf{G}}
  \widetilde{\mathbf{A}}_{\rm T}^{\mathsf{H}}$.

\section{Hybrid Limited Feedback Sparse Channel Estimation With Reduced Computational Cost}
\label{sec:hybrid_limited_feedback_sparse_channel_estimation}

The last setup proposed in this work is a hybrid between the setups presented in
Sections~\ref{sec:UE_based_limited_feedback_sparse_channel_estimation}
and~\ref{sec:Tx_based_limited_feedback_sparse_channel_estimation}.
This third setup is better suited to cases when the UE can afford to run simple channel estimation algorithms, such as OMP.
The  UE-based support identification algorithm presented in Algorithm~\ref{alg:OMP} is combined
with the BS-based limited feedback schemes of Section~\ref{sec:Tx_based_limited_feedback_sparse_channel_estimation}
resulting in an algorithm that can significantly reduce the computational cost at the BS, and possibly even the overall
feedback overhead for a given accuracy.

The UE first estimates the support of the downlink channel vector $\mathbf{g}$, $\mathcal{S}_{\widehat{\mathbf{g}}}$,
using   Algorithm~\ref{alg:OMP}. Let $\widehat{\mathbf{g}}$ be the $\overline{L}$-sparse channel estimate.%
\footnote{It is noted that having the support of complex vector $\widehat{\mathbf{g}}$
the support of $\widehat{\mathbf{x}}$ can be also inferred easily through Eq.~\eqref{eq:vector_x}.
Specifically, $\mathcal{S}_{\widehat{\mathbf{x}}} = \mathcal{S}_{\widehat{\mathbf{g}}} \cup \{G + \mathcal{S}_{\widehat{\mathbf{g}}}\} $.} %
 As feedback,  UE sends  the indices associated with non-zero elements of estimate $\widehat{\mathbf{g}}$  (i.e., $\mathcal{S}_{\widehat{\mathbf{g}}}$),
using $\overline{L}\, \mathsf{log}_2(G)$ bits, along with $2 N_{\rm fb}$ sign-quantized bits $\mathbf{b}$
associated with received signal $\mathbf{y}$.
  Upon receiving $\mathbf{b}$ and an estimate of the  support of $\mathbf{x}$,
the BS exploits the fact that the elements of  vector  $\widehat{\mathbf{x}}$ are zero  in the complement of
the support $\mathcal{S}_{\widehat{\mathbf{x}}}^{\mathtt{C}} = \{1,2,\ldots,2G\} \backslash \mathcal{S}_{\widehat{\mathbf{x}}}$, i.e.,
$\widehat{\mathbf{x}}_{\mathcal{S}_{\widehat{\mathbf{x}}}^{\mathtt{C}}} = \mathbf{0}$,   implying that
\begin{equation}
b_i =   \mathsf{sign}\!\left( \sum_{j \in \mathcal{S}_{\widehat{\mathbf{x}}}}
 {c}_{i,j} x_j +  z_i \right)  ,~i=1,2,\ldots,2N_{\rm fb},
\label{eq:model_BS_based_sparse}
\end{equation}
 and applies either of the two limited feedback channel estimation algorithms presented in
Sections~\ref{subsec:CS_formulation} and~\ref{subsec:sparse_ML_formulation},
but this time {\em limited to the reduced support}  $\mathcal{S}_{\widehat{\mathbf{x}}}$
to obtain an estimate  $\widehat{\mathbf{x}}_{\mathcal{S}_{\widehat{\mathbf{x}}}}$.
The whole procedure is listed in Algorithm~\ref{alg:HybridLimFeedAlgo}.

\begin{algorithm}[!t]
\caption{\small Hybrid Limited Feedback Sparse Channel Estimation}
 \begin{algorithmic} [1]
 \State{\small UE applies Algorithm~\ref{alg:OMP} to obtain support information $\mathcal{S}_{\widehat{\mathbf{g}}}$.}
  \State{\small UE  sends set $\mathcal{S}_{\widehat{\mathbf{g}}}$ and vector $\mathbf{b}$ using~\eqref{eq:real_equivalent1},
  requiring $ \overline{L}  \lceil\mathsf{log}_2 G\rceil + 2N_{\rm fb}$ feedback bits.}
\State{\small Upon receiving $\mathcal{S}_{\widehat{\mathbf{x}}}$  and $\mathbf{b}$, BS applies
Algorithm~\ref{alg:Limited_Feedback_Sparse_ML} or Eq.~\eqref{eq:CS_estimate_x}  to obtain
an estimate $\widehat{\mathbf{x}}_{\mathcal{S}_{\widehat{\mathbf{x}}}}$.}
\end{algorithmic}
\label{alg:HybridLimFeedAlgo}
\end{algorithm}

At the BS, the computational complexity of the proposed hybrid limited feedback sparse estimation
algorithms invoked in Algorithm~\ref{alg:HybridLimFeedAlgo}
is reduced by a factor $\overline{L}/G$ compared to the pure BS-based
counterparts of Section~\ref{sec:Tx_based_limited_feedback_sparse_channel_estimation}.
It is reasonable to assume that $\overline{L}$ is of the same order as $L$;
thus, using  extra $\lceil\overline{L}\, \mathsf{log}_2(G)\rceil$ feedback bits, the computational cost of BS reconstruction algorithms
executed over a reduced support
depends only on $N_{\rm fb}$ and $\overline{L}$ and becomes independent of the joint dictionary size $G$.
Numerical results show that not only the complexity diminishes, but the estimation error can
be further reduced compared to the case of not sending the support information.
This can in turn be used to reduce $N_{\rm fb}$, if so desired.

\section{Numerical Results}
\label{sec:numerical_results}

The double directional channel model in Eq.~\eqref{eq:dd_channel}
is used with uniform antenna directivity pattern at UE and uniform or 3GPP
antenna directivity pattern at the BS. BS and UE are equipped with ULAs.
A variety of performance metrics is examined such as normalized mean-squared error (NRMSE), beamforming gain,
and multiuser sum-capacity. The uplink feedback channel is considered error-free. The following algorithms are compared:
\begin{itemize}
\item LS channel estimation at the UE, given by  $\widehat{\mathbf{H}}_{\rm LS} = 
\mathbf{Y} \mathbf{S}^{\dagger}$,
 and quantization of    $\widehat{\mathbf{H}}_{\rm LS}$'s elements using scalar quantizer of $Q$ bits per real number. 
This feedback scheme requires exactly $2Q  M_{\rm T}   M_{\rm R}$ feedback bits.  This scheme is abbreviated LS-SQ.
\item For the case of $M_{\rm R}= 1$, we add in the comparisons a VQ technique that applies (a) LS channel estimation at the UE,
followed by (b) VQ of   $(\widehat{\mathbf{h}}_{\rm LS})^{\mathsf{H}}$, and (c) feedback of the VQ index. The VQ strategy of 
\cite{RyClVauCoGuoHon:09} based on a $2^Q$-PSK codebook $\mathcal{W}_{\rm PSK} 
\triangleq \left\{ \mathsf{e}^{\mathsf{j} 2 \pi \frac{(q-1)}{2^Q} } \right\}_{q=1}^{2^Q}$
is adopted for its good performance and low overhead ($Q  (M_{\rm T} -1)$ bits for channel feedback). This scheme is abbreviated LS-VQ.
\item    Combination of OMP   in Algorithm~\ref{alg:OMP}  with VQ  technique in~\cite{ChoLovMad:13}
% and  noncoherent trellis-coded quantization
using a rate 2/3 convolutional code. 
 The algorithm  exploits support information by executing first the OMP algorithm for support identification and then performs vector quantization over the reduced support.  
The number of states in the trellis diagram is $8$, and
parameter $Q$ determines the number of quantization phases in the optimization problem in \cite[Eq.~(12)]{ChoLovMad:13}, equal to $2^Q$.
The specific configuration for the algorithm 
in~\cite{ChoLovMad:13}    uses   $ \overline{L}(\lceil \mathsf{log}_2 G \rceil + 2) + 3$ feedback bits for the support information and the vector-quantized values, and is abbreviated OMP-VQ. 
\item The proposed UE-based baseline limited feedback scheme presented in Section~\ref{sec:UE_based_limited_feedback_sparse_channel_estimation}, henceforth  abbreviated OMP-SQ.
\item Single-bit CS limited feedback, as given in  Section~\ref{subsec:CS_formulation}.
\item Single-bit  $l_1$-regularized MLE limited feedback, as described in Section~\ref{subsec:sparse_ML_formulation}.
\item Hybrid single-bit $l_1$-regularized MLE and single-bit CS limited feedback
  algorithms, presented in Section~\ref{sec:hybrid_limited_feedback_sparse_channel_estimation}.
\end{itemize}
For scalar quantization,  LS-SQ uses Lloyd's algorithm for non-uniform quantization   and assumes perfect
knowledge of SQ thresholds at the BS.

\subsection{NRMSE vs. SNR}
\label{subsec:num_res_NRMSE_vs_SNR}

First   the impact of SNR on NRMSE performance for
the BS-based  schemes and hybrid counterparts is examined.
  NRMSE is defined as
$
  \mathbb{E}\!\left[  \frac{\| \widehat{\mathbf{H}} - \mathbf{H}\|_F}{ \| \mathbf{H} \|_F} \right].
$
The angle dictionary sizes for all algorithms were set to $G_{\rm T}= 140$
 and $G_{\rm R} = 16$. The number of scatterers, $L$, follows discrete uniform
 distribution over $[5,6,\ldots,9,10]$.
 We study two cases where  the azimuth angles $\phi_l$ and $\phi_l'$ (a) were drawn  uniformly from uniform angle dictionaries $\mathcal{P}_{\rm R}$ and $\mathcal{P}_{\rm T}$, both defined
 over $[-\pi/2, \pi/2)$; and (b) were random variables uniformly distributed over  $[-\pi/2, \pi/2)$,
i.e.,  $\phi_l, \phi_l' \sim \mathcal{U}[-\pi/2, \pi/2)$.
The remaining parameters were set as $M_{\rm R}=2$, $M_{\rm T} = 128$, $N_{\rm tr} = 64$, $N_{\rm fb} = N_{\rm tr} M_{\rm R} = 128$,
$P_{\rm T} = 1$ Watt. Rician fading was considered, i.e.,
$\alpha_l \sim \mathcal{CN}\!\left(  \sqrt{\frac{\kappa_l}{\kappa_l+1}}, \frac{1}{\kappa_l+1}\right)$,
 with $\kappa_l \sim  \mathcal{U}[0,40)$ and path delay $\varphi_l \sim \mathcal{U} [0,2\pi]$.
The dimensionality reducing matrix $\mathbf{P}$ for all algorithms
was a random selection of $N_{\rm fb}$ columns of the $ N_{\rm tr} M_{\rm R} \times N_{\rm fb}$   DFT matrix. %From Nikos: check if I understood correctly ...
Hybrid schemes use  $\overline{L}=15$.

 \begin{figure}[!t]
\centering
 \includegraphics[width=0.82\columnwidth]{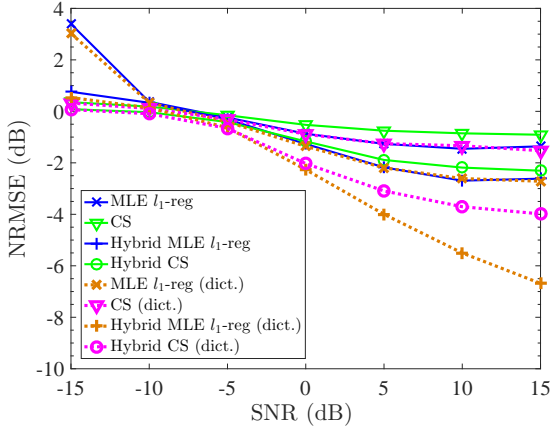} 
\caption{Comparison of NRMSE as a function of SNR for BS-based limited feedback schemes.}
\label{fig:NRMSE_vs_SNR_setup1_v1}
\end{figure}

Fig.~\ref{fig:NRMSE_vs_SNR_setup1_v1} shows the impact of  quantization error for AoA
and AoD. It can be seen that if the angles are drawn from the dictionaries there is
no  error due to angle quantization and the NRMSE of all studied algorithms becomes quite smaller
(brown and magenda dotted curves) than the case where the angles are drawn uniformly
in  $[-\pi/2, \pi/2)$  (green and blue solid curves).
The observation is that the impact of quantization error is severe for BS-based algorithms and their
hybrid counterparts, so to compensate for this, larger dictionary
sizes   $G_{\rm T}$ and  $G_{\rm R}$ will be utilized.
In what follows, we always draw $\phi_l, \phi_l' \sim \mathcal{U}[-\pi/2, \pi/2)$, so there is always dictionary mismatch.

\begin{figure}[!t]
\centering
         \includegraphics[width=0.81\columnwidth]{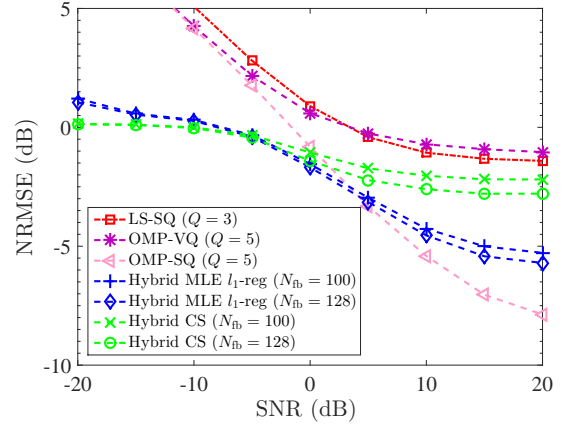}
\caption{The proposed methods use fewer feedback bits yet outperform the LS baseline for all values of SNR.}
\label{fig:NRMSE_vs_SNR_setup2_v1}
\end{figure}

Fig.~\ref{fig:NRMSE_vs_SNR_setup2_v1} compares  LS-SQ, OMP-VQ, OMP-SQ,  hybrid MLE $l_1$-regularized,  and  hybrid CS algorithms.
To alleviate quantization errors, the  proposed dictionary-based algorithms utilize $G_{\rm T} = G_{\rm R} = 240$.
Moreover we use $\overline{L}=15$ for the proposed limited feedback algorithms.
For fair comparison, we set parameters so that the number of feedback bits is of the same order of magnitude for all algorithms considered.
Note that for OMP-VQ the feedback overhead is not a function of $Q$
and thus it cannot be increased.
Hybrid $l_1$-regularized MLE and hybrid CS are executed with $N_{\rm fb} = 100$ and
$N_{\rm fb} = 120$, corresponding to  $440$ and $496$ feedback bits, respectively.
We set  $Q=3$ (corresponding to $1548$ feedback bits) for LS-SQ,  $Q = 5$ (corresponding to $390$ feedback bits)
for OMP-SQ, while OMP-VQ employs  $273$ feedback bits with $2^Q$  phase states.
%Note that it is not possible to make the number of bits exactly equal for all algorithms,
%but the numbers are at least comparable.

Fig.~\ref{fig:NRMSE_vs_SNR_setup2_v1} shows the NRMSE performance as a function of SNR. For SNR less than  $-5$ dB the hybrid
limited feedback schemes  achieve the best NRMSE performance. In the very low SNR regime the hybrid CS algorithm  offers the smallest NRMSE.
For SNR greater than $6$ dB, OMP-SQ with $Q=5$ outperforms the other algorithms.
%  The performance of  LS-SQ is the worst for all values of SNR.
The poor performance of LS-SQ  stems from the
fact that the soft estimate $\widehat{\mathbf{H}}_{\rm LS}$ before quantization is itself poor, as it does not exploit sparsity.
OMP-VQ offers the worst performance across all algorithms in the high SNR regime.
That happens because the VQ technique in  \cite{ChoLovMad:13} employs a predefined structured codebook at the BS, designed for Rayleigh fading.
Although the proposed algorithms use fewer feedback bits than LS-SQ, they achieve much better performance due to their judicious design.
For a moderate number of BS antennas, hybrid limited feedback algorithms are more suitable at low-SNR, while  OMP-SQ is better at high-SNR.

\subsection{NRMSE vs. $\overline{L}$}
\label{subsec:num_res_NRMSE_vs_Lbar}

\begin{figure}[!t]
\centering
        \includegraphics[width=0.82\columnwidth]{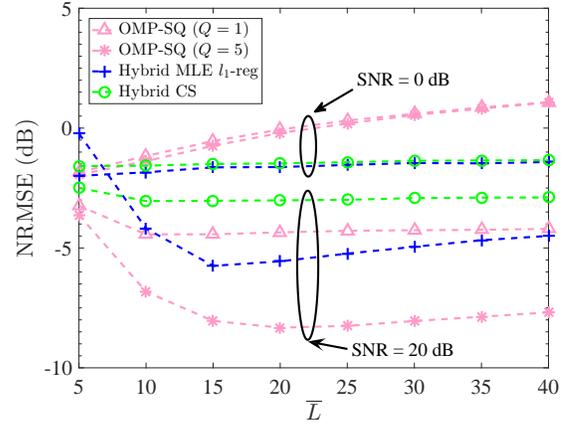}
\caption{NRMSE versus  $\overline{L}$ for the algorithms of setup 1 and 3. It can be seen that the NRMSE is not necessarily
a  decreasing function of $\overline{L}$.}
\label{fig:NRMSE_vs_Lbar_Mrx2_Mtx128_unif_pat}
\end{figure}

Using the same parameters as in the previous paragraph, Fig.~\ref{fig:NRMSE_vs_Lbar_Mrx2_Mtx128_unif_pat}
studies the impact of the maximum number of OMP iterations, $\overline{L}$, on NRMSE performance of OMP-SQ
and hybrid limited feedback algorithms under different values of SNR. It can be seen that in the low
SNR regime the smaller $\overline{L}$ is, the smaller NRMSE can be achieved
by all schemes. Namely, the smallest NRMSE is achieved for $\overline{L}=5$ for all algorithms.
On the other hand, in the high SNR regime the NRMSE versus $\overline{L}$ curve has a  convex
shape with minimum around $\overline{L} \in [10,20]$ for all algorithms. This indicates
that $\overline{L}$ should be chosen $\geq L$, but not much higher than $L$.

\subsection{NRMSE vs. $G$ and $M_{\rm T}$}
\label{subsec:num_res_NRMSE_vs_Q_VS_Mtx}

Next the impact of joint dictionary size, $G$, and the number of transmit antennas
on NRMSE is studied for the proposed  algorithms in Sections~\ref{sec:UE_based_limited_feedback_sparse_channel_estimation}
and~\ref{sec:hybrid_limited_feedback_sparse_channel_estimation}.
For this simulation, $M_{\rm R}=1$,
$N_{\rm tr} = 80$, and ${\rm SNR} =10$ dB.
Hybrid schemes utilize $N_{\rm fb} = 80 $,
the number of paths,  the maximum number of OMP iterations,  and  the dimensionality reducing matrix are the same as in Section~\ref{subsec:num_res_NRMSE_vs_SNR}.
OMP-SQ uses $Q=5$ bits per real number, and thus hybrid schemes and OMP-SQ utilize $160 + 15\mathsf{log}_2 G$
and $150 + 15\mathsf{log}_2 G$ feedback bits, respectively, in the worst case.

\begin{figure}[!t]
\centering
        \includegraphics[width=0.775\columnwidth]{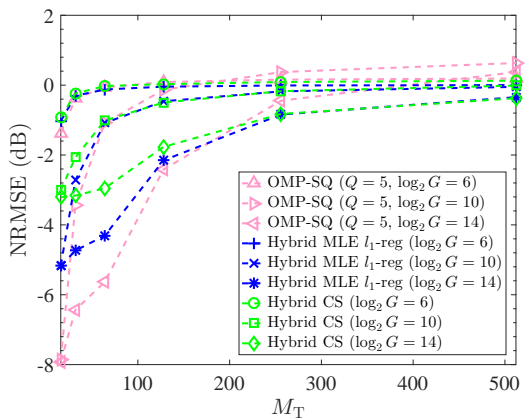}
\caption{NRMSE as a function of the number of BS antennas
for different joint dictionary sizes $G$. Higher $G$ improves  NRMSE.}
\label{fig:NRMSE_vsG_vsMtx_v1}
\end{figure}

 Fig.~\ref{fig:NRMSE_vsG_vsMtx_v1} studies the impact of
 $M_{\rm T}$ and $G$ on NRMSE. Recall that $G$ is determined
 from $G_{\rm T}$ and $G_{\rm R}$. Three different scenarios are considered
for $G$, using $G_{\rm T} = G_{\rm R} = 7$, $G_{\rm T} = G_{\rm R} = 31$,
and $G_{\rm T} = G_{\rm R} = 127$.
From Fig.~\ref{fig:NRMSE_vsG_vsMtx_v1} it is observed that for fixed
$G$ increasing the number of transmit antennas yields higher NRMSE,
while for fixed number of transmit   antennas, using higher $G$ yields reduced NRMSE, as expected.
Note that for $M_{\rm T} \geq 200$
OMP-SQ has the worst NRMSE performance, while for smaller $M_{\rm T}$ it achieves
better NRMSE compared to hybrid schemes.
For small $M_{\rm T}$,  increasing $G$ significantly reduces the NRSME. For large $M_{\rm T}$, increasing $G$ has little impact on the NRMSE.

\subsection{Beamforming Gain vs. SNR}
\label{subsec:num_res_BFG_vs_SNR}

\begin{figure}[!t]
\centering
\includegraphics[width=0.82\columnwidth]{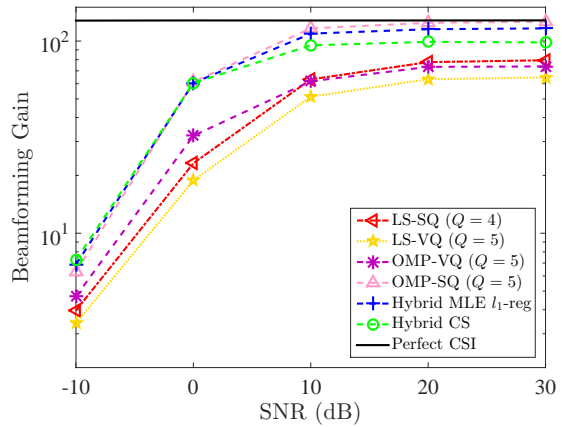}
\caption{Beamforming gain as a function of SNR for $5$ different algorithms.
The proposed methods outperform  LS schemes for all values of SNR.}
\label{fig:BFG_vs_SNR_Mrx1_Ntr80_G240_Mtx128_v3}
\end{figure}

Using the same   parameters as in Section~\ref{subsec:num_res_NRMSE_vs_SNR} (Fig.~\ref{fig:NRMSE_vs_SNR_setup2_v1})
with $M_{\rm T} = 128$, $M_{\rm R}=1$, and $P_{\rm T}=1$ Watt, in  Fig.~\ref{fig:BFG_vs_SNR_Mrx1_Ntr80_G240_Mtx128_v3} we study the beamforming gain  performance metric, defined as
$
 \mathbb{E} \left[ \frac{ P_{\rm T}   }  {   \|    \widehat{\mathbf{h}  }   \|_2^2 }   \left|
 \mathbf{h}^{\mathsf{H}}  \, \widehat{\mathbf{h}}\right |^2 \right]  ,
 \label{eq:avg_bf_gain_perf}
$
as a function of SNR. This metric measures the similarity between the actual channel $ \mathbf{h}$
and the normalized channel estimate $\widehat{\mathbf{h}}$ and is proportional to average received SNR.
We also include the performance of perfect CSI to assess an  upper bound
 on beamforming gain performance for the studied algorithms. %From Nikos: From Cauchy-Schwartz, the max beamforming gain (as defined) for transmit power equal to 1 as you wrote, will be the norm squared of the true channel. Is this equal to something over 100 here?
  Hybrid schemes of setup 3, OMP-VQ ($Q=5$),  OMP-SQ ($Q=5$), LS-SQ ($Q=4$), and LS-VQ ($Q=5$) use 
   $400$, $273$, $390$, $1024$, and $635$ 
feedback bits overhead, respectively. Interestingly, for ${\rm SNR} \geq 20$ dB  OMP-SQ with $Q=5$ achieves the performance of
 perfect CSI.  
 The proposed hybrid schemes have very similar but slightly worse performance relative to OMP-SQ.
%  However, the hybrid schemes use much fewer feedback bits.
 In addition, the performance gap between perfect CSI and the proposed algorithms
is less than $1.5$ dB for ${\rm SNR} \geq 10$ dB.   The proposed algorithms outperform LS schemes
for all values of SNR.  OMP-VQ performs very poorly compared to OMP-SQ
and other hybrid schemes.    OMP-SQ offers the best performance, but note that it {\em assumes}
perfect knowledge of the SQ thresholds at the BS, which in reality depend on the unknown channel.   
Perhaps surprisingly, LS-VQ offers smaller beamforming gain than LS-SQ.
One reason is that LS-SQ assumes perfect knowledge of the scalar quantization thresholds at the BS; another is that the vector-quantized codewords are confined to be PSK-codewords that lie on the
$M_{\rm T}$-dimensional unit complex circle, so magnitude variation among the elements of $\widehat{\mathbf{h}}_{\rm LS}^{\mathsf{H}}$ cannot be exploited. We also note that
the majority of VQ algorithms in the limited feedback literature, including LS-VQ, are
designed for  non-light-of-sight channels, a.k.a. Rayleigh fading, and the DD model used here is far from Rayleigh -- so LS-VQ
and OMP-VQ are not well-suited for the task.

\subsection{Beamforming Gain vs. $M_{\rm T}$}
\label{subsec:num_res_BFG_vs_Mtx}

A more realistic channel scenario is considered next,   based on the 3GPP multipath channel model \cite{3GPP_TR_36814_Rel9}, where path-loss and
shadowing are also incorporated in the path gains $\alpha_l$.
We assume a system operating at carrier frequency $F_{\rm car}= 2$ GHz,
and thus $\lambda \approx 0.15$. Transmit power and noise power are set
$0.5$ Watts and $10^{-10}$ Watts, respectively. The number of paths
is a discrete uniform RV in $ [5,6,\ldots,19,20] $.
For each path $l$:  $\phi_l, \phi_l' \sim \mathcal{U}[-\pi/2, \pi/2)$,
path distance $ d_l \sim \mathcal{U}[80, 120]$, common path-loss exponent    $ \eta \sim \mathcal{N}(2.8 , 0.1^2)$,
inverse path-loss        $\rho_l = \left(\frac{\lambda}{4 \pi}\right)^{2} \left( \frac{1}{d_l}\right)^{\eta}$,
shadowing  $ 10 \mathsf{log}_{10}(v_l)  \sim \mathcal{N}(  10 \mathsf{log}_{10}(\rho_l), 4^2) $,
and  Rician parameter $ \kappa_l \sim \mathcal{U}[0,50]$. Thus, the final
multipath gain is given by $ \alpha_l \sim \mathcal{CN}\!\left( \sqrt{\frac{\kappa_l}{\kappa_l+1} v_l}, \frac{1}{\kappa_l+1} v_l \right)$,
with  path delay $ \varphi_l \sim \mathcal{U}[0 , 2\pi]$.
The average received SNR, incorporating path-losses, small- and large- scale fading effects,
changes per realization, so  an implicit averaging with
respect to the received SNR is applied.
The  beamforming gain of all  algorithms compared in Section~\ref{sec:numerical_results}
is examined as a function of the number of transmit antennas.
 For this scenario we consider: $M_{\rm R} = 1$ received antenna, $N_{\rm tr} = 64$ training symbols,
$\overline{L} = 25$ for OMP and hybrid schemes, $N_{\rm fb} = N_{\rm tr}=64$ for all BS-based limited feedback algorithms and their hybrid counterparts,
and $N_{\rm fb}$ columns of the DFT matrix were chosen for the dimensionality reducing matrix $\mathbf{P}$.
The dictionary sizes  were set to $G_{\rm T} = G_{\rm R} = 180$.

 \begin{figure}[!t]
\centering
        \includegraphics[width=0.79\columnwidth]{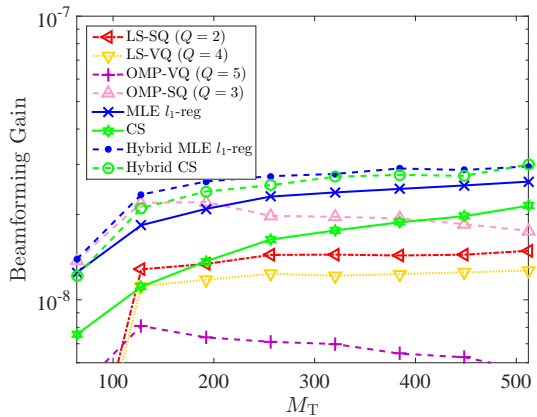}  
\caption{
  Beamforming gain as a function of $M_{\rm T}$ in the massive MIMO regime.
The proposed algorithms outperform LS-SQ and LS-VQ  for all values of $M_{\rm T}$.}
\label{fig:BFG_Mtx_64to512_G180_compare_all_unif_pattern_v3}
\end{figure}

Fig.~\ref{fig:BFG_Mtx_64to512_G180_compare_all_unif_pattern_v3}  examines a massive MIMO scenario
where $M_{\rm T}$ becomes very large. We observe that in this  scenario the beamforming gain takes values of order $10^{-8}$.
This is not surprising since on top of small-scale fading this scenario further incorporates path-loss and shadowing effects.
 
 From Fig.~\ref{fig:BFG_Mtx_64to512_G180_compare_all_unif_pattern_v3}  we note that hybrid $l_1$-regularized MLE achieves
 the best beamforming gain for almost all $M_{\rm T}$, while
hybrid CS has very similar performance. OMP-SQ and OMP-VQ are the only algorithms whose performance decreases as
 the number of transmit antennas increases.
  It should be noted that OMP-VQ ($Q=5$), OMP-SQ (with $Q = 3$), classic BS-based, and hybrid limited feedback schemes
utilize only $428,$ $524$, $128$, and $502$, feedback bits overhead, respectively.  
MLE $l_1$-reg and CS have worse performance than their hybrid counterparts.
On the other hand, LS-SQ (with $Q = 2$),  and LS-VQ (with $Q = 4$), employ  $4M_{\rm T}$
and $4  (M_{\rm T}-1)$ feedback bits overhead, that is linear in $M_{\rm T}$.
All proposed algorithms outperform  LS schemes as they exploit the inherent sparsity of the DD channel,
while the OMP-VQ algorithm offers very poor performance.  
It can be concluded that in the massive MIMO regime with realistic channel parameters, the BS-based limited feedback algorithms
 and their hybrid counterparts perform better than the other alternatives.

 \begin{figure}[!t]
\centering
        \includegraphics[width=0.82\columnwidth]{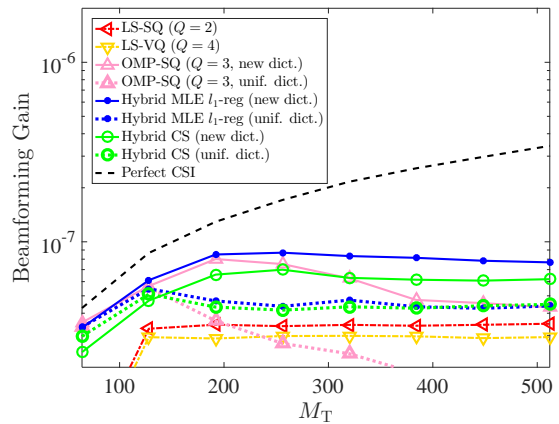}
\caption{Beamforming gain vs $M_{\rm T}$ using 3GPP antenna directivity pattern at the BS.
The  directivity pattern-aware dictionary outperforms uniform dictionary. }
\label{fig:BFG_vs_Mtx_3GPP_hybrid_schemes_patt_new_vs_unif_dict_v1}
\end{figure}

Next Fig.~\ref{fig:BFG_vs_Mtx_3GPP_hybrid_schemes_patt_new_vs_unif_dict_v1}  compares the proposed angle dictionary (labeled `new dict.') and the
uniform quantization dictionary (labeled `unif. dict.')
in the same massive MIMO scenario assuming that each BS antenna directivity pattern
is given by Eq.~\eqref{eq:ITU_antenna_pattern} using $\phi_{\rm 3dB} = 55^o$, $\mathtt{A}_{\rm m} = 30$ dB,
and $\mathtt{G}_{\rm dB} = 8$ dBi \cite{KamKhAltDebKam:14}.
All algorithms are configured with the same parameters as in the previous paragraph.
From Fig.~\ref{fig:BFG_vs_Mtx_3GPP_hybrid_schemes_patt_new_vs_unif_dict_v1}  is evident that for the same number of dictionary elements,
the proposed non-uniform directivity-based dictionary offers considerably higher beamforming gain performance compared to the uniform one.
In contrast to LS schemes, as the number of transmit antennas increases, the feedback overhead
for the proposed algorithms remains unaffected, rendering them a promising option for massive
MIMO systems.

\subsection{Execution Time vs. $M_{\rm T}$}
\label{subsec:num_res_exec_time_vs_Mtx}

\begin{figure}[!t]
\centering
\includegraphics[width=0.82\columnwidth]{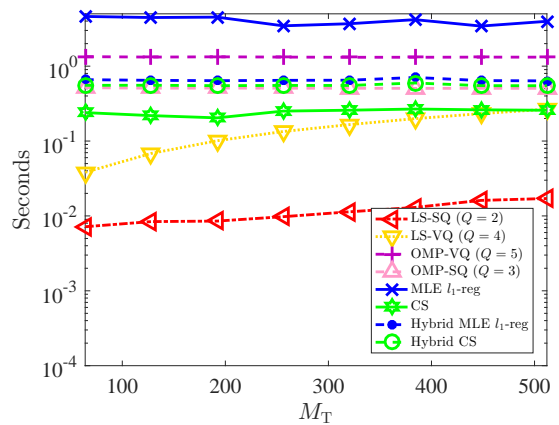}
\caption{Execution time  as a function of $M_{\rm T}$ in the massive MIMO regime.}
\label{fig:Execution_Time_vs_Mtx_Unif_patt_v1}
\end{figure}

For  the  pragmatic simulation setting of Fig.~\ref{fig:BFG_Mtx_64to512_G180_compare_all_unif_pattern_v3},
 Fig.~\ref{fig:Execution_Time_vs_Mtx_Unif_patt_v1} measures the end-to-end execution time of all algorithms averaged over 200 independent experiments. 
 It can be seen that MLE-reg algorithm
of setup 2 requires approximately 4 seconds for all values of $M_{\rm T}$, which is the highest
execution time. 
OMP-VQ algorithm requires approximately 1.3 seconds for all studied values of $M_{\rm T}$.
Hybrid schemes and OMP-SQ  offer end-to-end execution time of 0.5 seconds
for all values of $M_{\rm T}$, while CS scheme of setup 2 can reduce the execution time to the half, requiring 0.25 seconds.
As can be seen in Fig.~\ref{fig:Execution_Time_vs_Mtx_Unif_patt_v1}
  the execution time of the above algorithms remains unaffected by the number of BS antennas.
  On the contrary, the execution time of  the baseline LS schemes increases with the number of BS antennas.
The execution time of LS schemes is the smallest among all algorithms. When the number of 
BS antennas is moderate, LS-SQ and LS-VQ require execution time in the order of 0.01 seconds,
while in the massive MIMO regime their execution time increases to  0.017 and 0.25 seconds respectively.
The low execution time of  LS-SQ stems form the fact that it requires a calculation of a pseudoinverse
followed by the execution  of Lloyd's algorithm  using the build-in Matlab  functions.

\subsection{Multiuser Sum-Capacity vs. $P_{\rm T}$}
 \label{subsec:num_res_sumCap_vs_Mtx}

 In practice,  cellular systems serve concurrently multiple
 UE terminals at the same time, so a multiuser performance metric is of significant interest.
 Towards this end, we consider the downlink sum-capacity of a cellular
 network under zero-forcing ZF beamforming  as a function of $P_{\rm T}$, assuming $M_{\rm T}= 256$, $K=16$ scheduled UEs, $M_{\rm R} = 1$,
 and $N_{\rm tr} = 80$.
Hybrid schemes and OMP algorithms employ $\overline{L}$, $G_{\rm T} = 210$ and $G_{\rm R} = 180$
elements.
% The downlink spectral efficiency depends on the precoding  strategy utilized at the BS.
BS uses a data stream  of dimension $K$, $\mathbf{u} \in \mathds{C}^{K}$.
After receiving the feedback from $K$ UEs, BS  estimates the downlink channels  for each user
$k$, $\widehat{\mathbf{h}}_{k}^{\mathsf{H}}$,
 forms the compound downlink channel matrix
$
  \widehat{\mathbf{T}} =  \left[ \widehat{\mathbf{h}}_{1}~
  \widehat{\mathbf{h}}_{2}~\ldots~\widehat{\mathbf{h}}_{K}\right]^{\mathsf{H}}.
$
% and constructs the precoding matrix $\mathbf{V}$ from $ \widehat{\mathbf{T}}$ as
%$
%   \mathbf{V} =
%  [\mathbf{v}_1~\mathbf{v}_2~\ldots~\mathbf{v}_K].
%$
%BS transmits $\mathbf{s} = \mathbf{V} \mathbf{u}$.
%As in previous paragraph the precoding vector is normalized to
%satisfy the average total power constraint of $P_{\rm T}$.
Under ZF precoding with equal power allocation $\frac{P_{\rm T}}{K}$ per user, precoding matrix $\mathbf{V}$ is given by
$
\mathbf{V}= [\mathbf{v}_1~\mathbf{v}_2~\ldots~\mathbf{v}_K] =  \mathtt{t} \left(\widehat{\mathbf{T}}^{\mathsf{H}}
\widehat{\mathbf{T}}\right)^{-1} \widehat{\mathbf{T}}^{\mathsf{H}},
 \label{eq:ZF_precoding}
$
where $ \mathtt{t}^2 =  {\frac{K}{\mathsf{trace}\!\left(  \left(\widehat{\mathbf{T}}^{\mathsf{H}}
\widehat{\mathbf{T}}\right)^{-1}\right )}}$ guarantees that precoding vector satisfies the power constraint.
BS transmits $\mathbf{s} = \mathbf{V} \mathbf{u}$.
The corresponding  instantaneous signal-to-interference-plus-noise-ratio (SINR) for
user $k$ is given by
$
 \gamma_k = \frac{P_{\rm T}\left| \mathbf{h}_k^{\mathsf{H}} {\mathbf{v}}_{k} \right|^2}{ \sum \limits_{k' \neq k}
  P_{\rm T} \left|\mathbf{h}_k^{\mathsf{H}} {\mathbf{v}}_{k'} \right|^2 + K\, \sigma^2}.
$
The achievable ergodic rate for user $k$  is given by
$
\mathbb{E}[  \mathsf{R}(\gamma_k) ]=  \left(1 - \frac{N_{\rm tr }}{U_{\rm c}}\right)\mathbb{E}[ \mathsf{log}_2(1 +\gamma_k  )].
  \label{eq:rate_user_k}
$
% Denoting $\boldsymbol{\gamma} = [\gamma_1~\gamma_2~\ldots~\gamma_K]^{\top}$,
The achievable ergodic sum-rate (sum-capacity) for   $K$  scheduled UEs is expressed as
% \begin{equation}
% \mathbb{E}[  \mathsf{R}_{\rm sum}(\boldsymbol{\gamma})] =
$\sum_{k=1}^{K}  \mathbb{E}[ \mathsf{R}(\gamma_k)]$.
%   \label{eq:sum_rate}
% \end{equation}

  \begin{figure}[!t]
\centering
        \includegraphics[width=0.8\columnwidth]{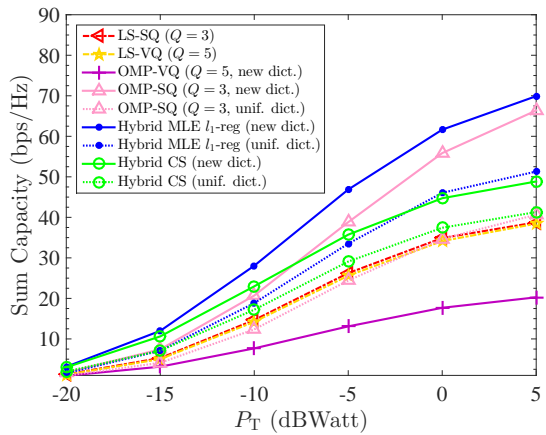}
\caption{Downlink sum-capacity as a function of the BS transmit power. The proposed
limited feedback along with non-uniform directional dictionaries schemes offer significant sum-capacity performance gains.}
\label{fig:capZF_vs_ptx_v1}
\end{figure}

\begin{table*}[!t]
% \vspace{-0.05 in}
\centering
\renewcommand{\arraystretch}{1.2}
\caption{Complexity analysis and number of feedback bits. }
 \begin{tabular}{ l|c|c|c|}
\centering
 &  Complexity at the BS &Complexity at the UE & Feedback Bits \\
\hline
LS-SQ  &  $\mathcal{O}(M_{\rm T}\, M_{\rm R})$  &    $\mathcal{O}\!\left(I_{\rm SQ} \, 2^{Q} \, M_{\rm T}\, M_{\rm R} + N_{\rm tr}^2(M_{\rm T} + N_{\rm tr})\right)$  & $2\,Q\,M_{\rm T}\,M_{\rm R}$  \\
\hline
LS-VQ  &  $\mathcal{O}(1)$   &    $\mathcal{O}\!\left( M_{\rm T} \, \mathsf{log}( M_{\rm T})+ N_{\rm tr}^2(M_{\rm T} + N_{\rm tr})\right)$ & $ Q \, (M_{\rm T}-1)$ \\
\hline
OMP-SQ &   $\mathcal{O}(\overline{L} \, M_{\rm T}\, M_{\rm R})$  &
   $\mathcal{O}\!\left(I_{\rm SQ} \, 2^{Q}\, \overline{L} + \overline{L}  \, M_{\rm R} \,N_{\rm tr} (\overline{L}  + G)\right)$ & $   \overline{L} ( \lceil\mathsf{log}_2 G\rceil   + 2Q)$\\
\hline
OMP-VQ &  $\mathcal{O}(\overline{L} ( M_{\rm T}\, M_{\rm R} + 2^{ 5+Q}))$   &
  $\mathcal{O}\!\left(2^{ 5+Q}\, \overline{L} + \overline{L}  \, M_{\rm R} \,N_{\rm tr} (\overline{L}  + G)\right)$   &$  \overline{L}(\lceil \mathsf{log}_2 G \rceil + 2) + 3 $  \\
\hline
CS &  $\mathcal{O}\!\left(  G (N_{\rm fb} + M_{\rm T}\, M_{\rm R})  \right) $ &   $N_{\rm fb} \, M_{\rm R} \, N_{\rm tr}$  &  $2 N_{\rm fb} $\\
\hline
MLE-reg &   $\mathcal{O}\!\left( G( (I_{\rm max} + N_{\rm fb})  N_{\rm fb} + M_{\rm T}\, M_{\rm R})   \right) $ &  $N_{\rm fb} \, M_{\rm R} \, N_{\rm tr}$  &  $  2  N_{\rm fb}$ \\
\hline
Hybrid CS &   $\mathcal{O}\!\left(\overline{L}(N_{\rm fb} +  M_{\rm T}\, M_{\rm R}) \right)$ &  $\mathcal{O}\!\left(N_{\rm fb} \, M_{\rm R} \, N_{\rm tr} + \overline{L}  \, M_{\rm R} \,N_{\rm tr} (\overline{L}  + G)\right)$  &  $   2  N_{\rm fb} + \overline{L}\lceil\mathsf{log}_2 G\rceil  $\\
\hline
Hybrid MLE-reg &  $\mathcal{O}\!\left(\overline{L}((I_{\rm max} + N_{\rm fb})   N_{\rm fb} +  M_{\rm T}\, M_{\rm R}) \right)$ & $\mathcal{O}\!\left(N_{\rm fb} \, M_{\rm R} \, N_{\rm tr} + \overline{L}  \, M_{\rm R} \,N_{\rm tr} (\overline{L}  + G)\right)$   &  $ 2  N_{\rm fb} + \overline{L}\lceil\mathsf{log}_2 G\rceil  $ \\
\hline
 \end{tabular}
\label{table:complexity}
 \end{table*}

Fig.~\ref{fig:capZF_vs_ptx_v1} depicts the downlink sum-capacity
% in~\eqref{eq:sum_rate}
as a function of  BS transmit power $P_{\rm T}$.
The downlink channels for each user are generated using the same  parameters as in Section \ref{subsec:num_res_BFG_vs_Mtx}
with antenna directivity pattern parameters    $\phi_{\rm 3dB} = 55^o$, $\mathtt{A}_{\rm m} = 30$ dB,
and $\mathtt{G}_{\rm dB} = 8$ dBi.
The coherence block occupies 20 resource blocks,  i.e., $U_{\rm c} = 1680$ channel uses.
The following algorithms are compared: LS-SQ with $Q=3$, LS-VQ with $Q=5$, OMP-VQ with $Q=5$, OMP-SQ with $Q=3$,  and
hybrid schemes using the proposed dictionaries (labeled `new dict.') and uniform dictionaries (labeled `unif. dict.').
The performance gains of the proposed non-uniform dictionaries over conventional uniform ones
are evident in Fig.~\ref{fig:capZF_vs_ptx_v1}, especially for MLE-reg and OMP-SQ algorithms.
For $1$ Watt transmission power, MLE-reg and OMP-SQ with  proposed non-uniform dictionaries
offer $15$ and $20$ bit/sec/Hz higher capacity than  MLE-reg and OMP-SQ executed with uniform dictionaries.
The proposed methods in conjunction with non-uniform dictionaries offer a substantial sum-capacity performance gain  compared to LS schemes.
The performance of OMP-VQ is very poor, at least $5$ dB worse than proposed MLE-reg algorithm with non-uniform dictionaries
for all values of $P_{\rm T}$.

\subsection{Complexity Analysis}
 \label{subsec:complexity_analysis}

In this section a detailed  computational  complexity analysis at both UE and BS is presented
 for all studied algorithms. Table~\ref{table:complexity} shows the computational cost of all studied
algorithms along with the required number of feedback bits.

For LS schemes, at the UE side the calculation of $\widehat{\mathbf{H}}_{\rm LS} $ requires
$\mathcal{O}( N_{\rm tr}^2(M_{\rm T} + N_{\rm tr}))$ arithmetic operations. For LS-SQ, at the UE side,
for each element  of $\widehat{\mathbf{H}}_{\rm LS}$,     $\mathcal{O}(I_{\rm SQ}\, 2^{Q})$ computations
are required for the SQ algorithm,
where $I_{\rm SQ}$ is the maximum of iterations for algorithm to  converge.
After receiving the associated indices and the elements of the quantized channel, the BS
reconstructs the channel with complexity  $\mathcal{O}(M_{\rm T}\, M_{\rm R})$.
For LS-VQ, at the UE side, the computational cost is due to the calculation of $\widehat{\mathbf{H}}_{\rm LS} $
and the computation of optimal $M_{\rm T}$-dimensional $2^{Q}$-PSK sequence, which requires
$\mathcal{O}( M_{\rm T} \, \mathsf{log}( M_{\rm T}))$ computations  \cite{RyClVauCoGuoHon:09}. Since the codebook is
already stored at the BS the channel reconstruction requires $\mathcal{O}(1)$ computations.  
The complexity  of OMP algorithm is dominated by lines 4, 7, and 8
in Algorithm~\ref{alg:OMP}, which is  $\overline{L}  \, M_{\rm R} \,N_{\rm tr} (\overline{L}  + G) )$.
Hence, the complexity for OMP-SQ is  $\mathcal{O}\!\left(I_{\rm SQ} \, 2^{Q}\, \overline{L} +
\overline{L}  \, M_{\rm R} \,N_{\rm tr} (\overline{L}  + G)\right)$.
At the BS, the reconstruction of the channel matrix for OMP-SQ exploits the sparsity of
channel vector $\widehat{\mathbf{g}}$, and thus using only  the $\overline{L}$
non-zero elements of sparse matrix $\widehat{\mathbf{G}}$ the channel reconstruction
using~\eqref{eq:dd_channel_model_dictionary_compact1} requires only
$\mathcal{O}(\overline{L} \, M_{\rm T}\, M_{\rm R})$ arithmetic operations.
The complexity of  the algorithm in  \cite{ChoLovMad:13} at the UE side is due to the vector quantization of  the  $\overline{L}$
non-zero elements of path coefficient vector $\widehat{\mathbf{g}}$ through Viterbi algorithm
 ($ 2^{ 5+Q} \overline{L} $ operations) and the
support identification of  $\widehat{\mathbf{g}}$ through OMP algorithm. Hence, OMP-VQ algorithm requires
total  $\mathcal{O}\!\left(2^{ 5+Q}\, \overline{L} + \overline{L}  \, M_{\rm R} \,N_{\rm tr} (\overline{L}  + G)\right)$
arithmetic operations at the UE.
At the BS side, OMP-VQ algorithm reconstructs
the non-zero elements of $\widehat{\mathbf{g}}$ through the Viterbi algorithm,
requiring $O(2^{ 5+Q}\, \overline{L})$ computations; whereas the reconstruction of
the actual channel, using~\eqref{eq:dd_channel_model_dictionary_compact1},
demands  $\mathcal{O}(\overline{L} \, M_{\rm T}\, M_{\rm R})$ arithmetic operations. 
BS-based limited feedback schemes require $ N_{\rm fb} \, M_{\rm R} \, N_{\rm tr}$
  arithmetic operations at the UE side due to the multiplication of $\mathbf{P}^{\mathsf{H}}$
with $\mathbf{y}$. While hybrid schemes require an extra  $\overline{L}  \, M_{\rm R} \,N_{\rm tr} (\overline{L}  + G) )$
computational cost at the UE side due to the execution of OMP algorithm for support identification. 
At the BS side, as shown in   Sections~\ref{subsec:CS_formulation} and~\ref{subsec:sparse_ML_formulation},
 CS and MLE-reg algorithms require    $\mathcal{O}(  G  \, N_{\rm fb} )$  and  $\mathcal{O}( (I_{\rm max} + N_{\rm fb})  G \, N_{\rm fb} )$ computations,
respectively, to obtain an estimate of vector $\mathbf{x}$. In addition, an extra  $\mathcal{O}(G \, M_{\rm T}\, M_{\rm R})$
computational cost is required to reconstruct the actual channel through~\eqref{eq:dd_channel_model_dictionary_compact1}.
Finally, hybrid schemes require at the BS,  $\mathcal{O}(  \overline{L}  \, N_{\rm fb} )$  for CS and  $\mathcal{O}( (I_{\rm max} + N_{\rm fb})   \overline{L} \, N_{\rm fb} )$
for MLE-reg algorithms. Using the support information obtained from feedback, hybrid schemes require   extra  $\mathcal{O}(\overline{L} M_{\rm T}\, M_{\rm R})$
calculations to evaluate~\eqref{eq:dd_channel_model_dictionary_compact1} for channel reconstruction.

\subsection{Take-home Points from the Simulations}
 \label{subsec:remarks}
We close this section by summarizing the most important take-home points from our numerical results. 
 
The baseline quantization algorithms LS-SQ and LS-VQ   require low execution time but their
 feedback overhead   scales linearly with the number of BS antennas. As they don't 
exploit the DD parameterization, LS-SQ and LS-VQ yield worse estimation accuracy compared to the proposed limited feedback algorithms of setups 1, 2, and 3, even though LS-SQ/VQ use a higher number of feedback bits. 
 OMP-VQ requires relative   execution time and   feedback overhead, depending   linearly on $\overline{L}$
 and logarithmically on $G$;
it performs very poorly in all our simulation scenarios. The principal reason for the poor performance for OMP-VQ is that its codewords 
are pre-defined and fixed, offering limited  channel estimation granularity. The proposed algorithms of setup  2 require 
$2 N_{\rm fb} \leq 2 N_{\rm tr} M_{\rm R}$ feedback bits, independent of the number of BS antennas. 
The execution time of one-bit MLE-reg is high, whereas the execution time of one-bit CS is at least an order of magnitude lower; 
on the other hand,   the estimation  performance of 
one-bit CS is   worse compared to one-bit MLE-reg. Both algorithms of
setup 2 have slightly worse  % \reminder{worse or better? I guess you mean one-bit CS is slightly worse, 
%while one-bit MLE-reg is slightly better. So you have to write this as "The performance of one-bit CS (one-bit MLE-reg) is slightly worse (resp. better) ..."} 
estimation performance than OMP-SQ for moderate number of BS antennas in the high   SNR regime.
Conversely, in the low SNR regime or when the number of BS antennas increases, one-bit CS and one-bit MLE-reg outperform OMP-SQ.
% while in the massive MIMO regime, they outperform OMP-SQ.
Moreover, the algorithms of setup 2 perform slightly worse compared to their hybrid counterparts of setup 3.
Such performance gains of hybrid schemes come at the cost of an extra
 $\overline{L} \lceil \mathsf{log}_2 G \rceil$ bits in feedback overhead.

 We found that 
employing joint dictionary size   $G = \mathcal{O}(M_{\rm T} M_{\rm R})$  suffices to obtain 
good channel estimation accuracy for the dictionary-based algorithms, 
corroborating the findings of dictionary-based estimation  
 prior art \cite{BaHaSaNo:10,  HeGoRaRoSay:16, MerRusGonAlkHe:16}.
Consequently,  the feedback overhead for massive MIMO systems
employing the proposed OMP-SQ and hybrid schemes, scales as $\mathcal{O}( \overline{L} \, 
( \mathsf{log} ( M_{\rm T}M_{\rm R}) + 2Q) )$ and 
  $\mathcal{O}( \overline{L} \,   \mathsf{log}( M_{\rm T}M_{\rm R})  +  2N_{\rm fb})$, respectively. 
This logarithmic scaling with the number of BS antennas underscores the practicality of the proposed limited feedback algorithms for the DD model in massive MIMO scenarios.

\section{Conclusion and Future Work}
\label{sec:conclusion}

This work provided a new limited feedback
framework using dictionary-based sparse channel estimation algorithms that
entail low computational complexity, and thus can be implemented in  real-time.
The proposed dictionary accounts for the antenna directivity pattern and can offer
beamforming and capacity gains while requiring less feedback overhead compared to uniform dictionaries.
Unlike VQ-based schemes for which the number of feedback bits must grows linearly with the
number of BS antennas to maintain a certain performance level, the  number of feedback bits for the proposed algorithms
is under designer control, and they can achieve better performance using a substantially lower bit budget.
The proposed baseline OMP-SQ algorithm (setup 1) achieves the best performance when the number of
transmit antennas is moderate and SNR is high, while in the low-SNR regime the BS-based (setup 2) and hybrid (setup 3) schemes  offer
better performance. The hybrid schemes (setup 3) achieve the best performance in the massive MIMO regime.

%\bibliographystyle{IEEEtran}
%\bibliography{IEEEabrv,bibtex_channel_estimation}

\begin{thebibliography}{10}
\providecommand{\url}[1]{#1}
\csname url@samestyle\endcsname
\providecommand{\newblock}{\relax}
\providecommand{\bibinfo}[2]{#2}
\providecommand{\BIBentrySTDinterwordspacing}{\spaceskip=0pt\relax}
\providecommand{\BIBentryALTinterwordstretchfactor}{4}
\providecommand{\BIBentryALTinterwordspacing}{\spaceskip=\fontdimen2\font plus
\BIBentryALTinterwordstretchfactor\fontdimen3\font minus
  \fontdimen4\font\relax}
\providecommand{\BIBforeignlanguage}[2]{{%
\expandafter\ifx\csname l@#1\endcsname\relax
\typeout{** WARNING: IEEEtran.bst: No hyphenation pattern has been}%
\typeout{** loaded for the language `#1'. Using the pattern for}%
\typeout{** the default language instead.}%
\else
\language=\csname l@#1\endcsname
\fi
#2}}
\providecommand{\BIBdecl}{\relax}
\BIBdecl

\bibitem{HoyBriDeb:13}
J.~Hoydis, S.~ten Brink, and M.~Debbah, ``Massive {MIMO} in the {UL/DL} of
  cellular networks: How many antennas do we need?'' \emph{IEEE J. Sel. Areas
  Commun.}, vol.~31, no.~2, pp. 160--171, Feb. 2013.

\bibitem{Mar:10}
T.~L. Marzetta, ``Noncooperative cellular wireless with unlimited numbers of
  base station antennas,'' \emph{IEEE Trans. Wireless. Comm.}, vol.~9, no.~11,
  pp. 3590--3600, Nov. 2010.

\bibitem{RuPeLauLarMaeEdTuf:13}
F.~Rusek \emph{et~al.}, ``Scaling up {MIMO}: Opportunities and challenges with
  very large arrays,'' \emph{IEEE Signal Process. Mag.}, vol.~30, no.~1, pp.
  40--60, Jan. 2013.

\bibitem{NgLarMar:13}
H.~Q. Ngo, E.~G. Larsson, and T.~L. Marzetta, ``Energy and spectral efficiency
  of very large multiuser {MIMO} systems,'' \emph{IEEE Trans. Commun.},
  vol.~61, no.~4, pp. 1436--1449, Apr. 2013.

\bibitem{Jin:06}
N.~Jindal, ``{MIMO} broadcast channels with finite-rate feedback,'' \emph{IEEE
  Trans. Inf. Theor.}, vol.~52, no.~11, pp. 5045--5060, Nov. 2006.

\bibitem{CaJiKoRa:10}
G.~Caire, N.~Jindal, M.~Kobayashi, and N.~Ravindran, ``Multiuser {MIMO}
  achievable rates with downlink training and channel state feedback,''
  \emph{IEEE Trans. Inf. Theor.}, vol.~56, no.~6, pp. 2845--2866, Jun. 2010.

\bibitem{AdNamAhnCai:13}
A.~Adhikary, J.~Nam, J.-Y. Ahn, and G.~Caire, ``Joint spatial division and
  multiplexingâ€”the large-scale array regime,'' \emph{IEEE Trans. Inf.
  Theor.}, vol.~59, no.~10, pp. 6441--6463, Oct. 2013.

\bibitem{DahParkSk:16}
E.~Dahlman, S.~Parkvall, and J.~Skold, \emph{4G, {LTE}-{A}dvanced {P}ro and The
  Road to {5G}}.\hskip 1em plus 0.5em minus 0.4em\relax Elsevier Science, 2016.

\bibitem{Hyo_et_all:16}
H.~Ji \emph{et~al.}, ``Overview of full-dimension {MIMO} in {LTE-A}dvanced
  {P}ro.'' \emph{CoRR}, 2016.

\bibitem{Love_et_al:08}
D.~J. Love, R.~W. Heath, V.~K. Lau, D.~Gesbert, B.~D. Rao, and M.~Andrews, ``An
  overview of limited feedback in wireless communication systems,''
  \emph{{IEEE} J. Sel. Areas Commun.}, vol.~26, no.~8, pp. 1341--1365, Oct.
  2008.

\bibitem{MuSabErkAaz:03}
K.~K. Mukkavilli, A.~Sabharwal, E.~Erkip, and B.~Aazhang, ``On beamforming with
  finite rate feedback in multiple-antenna systems,'' \emph{IEEE Trans. Inf.
  Theor.}, vol.~49, no.~10, pp. 2562--2579, Oct. 2003.

\bibitem{LaLiCh:04}
V.~Lau, Y.~Liu, and T.-A. Chen, ``On the design of {MIMO} block-fading channels
  with feedback-link capacity constraint,'' \emph{{IEEE} Trans. Commun.},
  vol.~52, no.~1, pp. 62--70, Jan. 2004.

\bibitem{RyClVauCoGuoHon:09}
D.~J. Ryan, I.~V.~L. Clarkson, I.~B. Collings, D.~Guo, and M.~L. Honig, ``{QAM}
  and {PSK} codebooks for limited feedback {MIMO} beamforming,'' \emph{{IEEE}
  Trans. Commun.}, vol.~57, no.~4, pp. 1184--1196, Apr. 2009.

\bibitem{ChoLovMad:13}
J.~Choi, Z.~Chance, D.~J. Love, and U.~Madhow, ``Noncoherent trellis coded
  quantization: A practical limited feedback technique for massive {MIMO}
  systems,'' \emph{{IEEE} Trans. Commun.}, vol.~61, no.~12, pp. 5016--5029,
  Dec. 2013.

\bibitem{XiaGian:06}
P.~Xia and G.~Giannakis, ``Design and analysis of transmit-beamforming based on
  limited-rate feedback,'' \emph{IEEE Trans. Signal Process.}, vol.~54, no.~5,
  pp. 1853--1863, May 2006.

\bibitem{HuHeAn:09}
K.~Huang, R.~W. Heath~Jr, and J.~G. Andrews, ``Limited feedback beamforming
  over temporally-correlated channels,'' \emph{IEEE Trans. Signal Process.},
  vol.~57, no.~5, pp. 1959--1975, May 2009.

\bibitem{MehSid:14_a}
O.~Mehanna and N.~D. Sidiropoulos, ``Channel tracking and transmit beamforming
  with frugal feedback,'' \emph{IEEE Trans. Signal Process.}, vol.~62, no.~24,
  pp. 6402--6413, Dec. 2014.

\bibitem{MarHoch:06}
T.~L. Marzetta and B.~M. Hochwald, ``Fast transfer of channel state information
  in wireless systems.'' \emph{IEEE Trans. Signal Process.}, vol.~54, no.~4,
  pp. 1268--1278, Apr. 2006.

\bibitem{JiaMoCaiZhi:15}
Z.~Jiang, A.~F. Molisch, G.~Caire, and Z.~Niu, ``Achievable rates of {FDD}
  massive {MIMO} systems with spatial channel correlation,'' \emph{IEEE Trans.
  Wireless. Comm.}, vol.~14, no.~5, pp. 2868--2882, May 2015.

\bibitem{BaHaSaNo:10}
W.~U. Bajwa, J.~Haupt, A.~M. Sayeed, and R.~Nowak, ``Compressed channel
  sensing: A new approach to estimating sparse multipath channels,''
  \emph{Proc. IEEE}, vol.~98, no.~6, pp. 1058--1076, Jun. 2010.

\bibitem{HeGoRaRoSay:16}
R.~W. Heath, N.~Gonzalez-Prelcic, S.~Rangan, W.~Roh, and A.~M. Sayeed, ``An
  overview of signal processing techniques for millimeter wave {MIMO}
  systems,'' \emph{{IEEE} J. Sel. Topics Signal Process.}, vol.~10, no.~3, pp.
  436--453, May 2016.

\bibitem{3GPP_TS_36101_Rel13}
$\text{3GPP TS 36.101 V13.2.1}$, ``{E}volved universal terrestrial radio access
  ({E-UTRA}); {U}ser {E}quipment ({UE}) radio transmission and reception,
  {R}elease 13,'' May 2016.

\bibitem{KamKhAltDebKam:14}
A.~Kammoun, H.~Khanfir, Z.~Altman, M.~Debbah, and M.~Kamoun, ``Preliminary
  results on {3D} channel modeling: From theory to standardization,''
  \emph{{IEEE} J. Sel. Areas Commun.}, vol.~32, no.~6, pp. 1219--1229, Jun.
  2014.

\bibitem{MerRusGonAlkHe:16}
R.~M{\'e}ndez-Rial, C.~Rusu, N.~Gonz{\'a}lez-Prelcic, A.~Alkhateeb, and R.~W.
  Heath, ``Hybrid {MIMO} architectures for millimeter wave communications:
  Phase shifters or switches?'' \emph{IEEE Access}, vol.~4, pp. 247--267, Jan.
  2016.

\bibitem{MoShPrHe:14}
J.~Mo, P.~Schniter, N.~G. Prelcic, and R.~W. Heath, ``Channel estimation in
  millimeter wave {MIMO} systems with one-bit quantization,'' in \emph{Proc.
  Asilomar Conf. on Signals, Systems and Computers (Asilomar)}, Pacific Grove,
  CA, 2014, pp. 957--961.

\bibitem{RuRiPrHe:15}
C.~Rusu, R.~M{\'e}ndez-Rial, N.~Gonz{\'a}lez-Prelcic, and J.~R.~W.~Heath,
  ``Adaptive one-bit compressive sensing with application to low-precision
  receivers at mm{W}ave,'' in \emph{Proc. IEEE Global Telecommunications Conf.
  (GLOBECOM)}, San Diego, CA, Dec. 2015.

\bibitem{ZhChGuHo:17}
Z.~Zhou, X.~Chen, D.~Guo, and M.~L. Honig, ``Sparse channel estimation for
  massive {MIMO} with 1-bit feedback per dimension,'' in \emph{Proc. IEEE
  {W}ireless {C}ommun. and {N}etworking {C}onf. (WCNC)}, San Francisco, CA,
  2017.

\bibitem{TroGil:07}
J.~A. Tropp and A.~C. Gilbert, ``Signal recovery from random measurements via
  orthogonal matching pursuit,'' \emph{IEEE Trans. Inf. Theor.}, vol.~53,
  no.~12, pp. 4655--4666, Dec. 2007.

\bibitem{ChDiRaKi:15}
J.~W. {Choi}, B.~{Shim}, Y.~{Ding}, B.~{Rao}, and D.~{In Kim}, ``Compressed
  sensing for wireless communications : Useful tips and tricks,'' \emph{ArXiv
  e-prints}, Nov. 2015.

\bibitem{Nes:04}
Y.~Nesterov, \emph{Introductory lectures on convex optimization : a basic
  course}, ser. Applied optimization.\hskip 1em plus 0.5em minus 0.4em\relax
  Boston, Dordrecht, London: Kluwer Academic Publ., 2004.

\bibitem{ODonCan:15}
B.~O'Donoghue and E.~Cand{\`e}s, ``Adaptive restart for accelerated gradient
  schemes,'' \emph{Foundations of Computational Mathematics}, vol.~15, no.~3,
  pp. 715--732, 2015.

\bibitem{GrNeu:99}
R.~M. Gray and D.~L. Neuhoff, ``Quantization,'' \emph{IEEE Trans. Inf. Theor.},
  vol.~44, no.~6, pp. 2325--2383, Oct. 1998.

\bibitem{3GPP_TR_36814_Rel9}
$\text{3GPP TS 36.814 V9.0.0}$, ``{E}volved universal terrestrial radio access
  ({E-UTRA}); {F}urther advancements for {E-UTRA} physical layer aspects,
  {R}elease 9,'' Mar. 2010.

\bibitem{AlFuSidYaBl:17}
P.~N. Alevizos, X.~Fu, N.~Sidiropoulos, Y.~Yang, and A.~Bletsas, ``Non-uniform
  directional dictionary-based limited feedback for massive {MIMO} systems,''
  in \emph{Proc. IEEE International Symposium on Modeling and Optimization in
  Mobile, Ad Hoc, and Wireless Networks (WiOpt)}, Paris, FR, May 2017.

\bibitem{SeTaBa:11}
\BIBentryALTinterwordspacing
S.~Sesia, I.~Toufik, and M.~Baker, \emph{{LTE} - The {UMTS} Long Term
  Evolution: From Theory to Practice}.\hskip 1em plus 0.5em minus 0.4em\relax
  Wiley, 2011. [Online]. Available:
  \url{https://books.google.es/books?id=beIaPXLzYKcC}
\BIBentrySTDinterwordspacing

\bibitem{3GPP_TS_37840_Rel12}
$\text{3GPP TR 37.840 V12.1.0}$, ``{T}echnical {S}pecification {G}roup {R}adio
  {A}ccess {N}etwork; {S}tudy of {R}adio {F}requency ({RF}) and
  {E}lectromagnetic {C}ompatibility ({EMC}) requirements for {A}ctive {A}ntenna
  {A}rray {S}ystem ({AAS}) base station, {R}elease 12,'' Dec. 2013.

\bibitem{Blum:12}
T.~Blumensath, ``Accelerated iterative hard thresholding,'' \emph{Signal
  Processing}, vol.~92, no.~3, pp. 752--756, 2012.

\bibitem{FoucRau:13}
S.~Foucart and H.~Rauhut, \emph{A Mathematical Introduction to Compressive
  Sensing}.\hskip 1em plus 0.5em minus 0.4em\relax Birkh\"{a}user Basel, 2013.

\bibitem{BouBar:08}
P.~T. Boufounos and R.~G. Baraniuk, ``1-bit compressive sensing,'' in
  \emph{Proc. IEEE Information Sciences and Systems (CISS)}, 2008, pp. 16--21.

\bibitem{PlanVer:13_a}
Y.~Plan and R.~Vershynin, ``One-bit compressed sensing by linear programming,''
  \emph{Communications on Pure and Applied Mathematics}, vol.~66, no.~8, pp.
  1275--1297, 2013.

\bibitem{JacLasBouBar:13}
L.~Jacques, J.~N. Laska, P.~T. Boufounos, and R.~G. Baraniuk, ``Robust 1-bit
  compressive sensing via binary stable embeddings of sparse vectors,''
  \emph{IEEE Trans. Inf. Theor.}, vol.~59, no.~4, pp. 2082--2102, Apr. 2013.

\bibitem{ZhaYiJin:14}
L.~Zhang, J.~Yi, and R.~Jin, ``Efficient algorithms for robust one-bit
  compressive sensing,'' in \emph{Proc. International Conference on Machine
  Learning (ICML)}, Beijing, China, Jun. 2014, pp. 820--828.

\bibitem{TsaJaSidOtt:13}
E.~Tsakonas, J.~Jald{\'e}n, N.~D. Sidiropoulos, and B.~Ottersten, ``Sparse
  conjoint analysis through maximum likelihood estimation,'' \emph{IEEE Trans.
  Signal Process.}, vol.~61, no.~22, pp. 5704--5715, Nov. 2013.

\bibitem{BoydVand:04}
S.~Boyd and L.~Vandenberghe, \emph{Convex Optimization}.\hskip 1em plus 0.5em
  minus 0.4em\relax New York, NY, USA: Cambridge University Press, 2004.

\bibitem{Bert:15}
D.~P. Bertsekas, \emph{Convex optimization algorithms}.\hskip 1em plus 0.5em
  minus 0.4em\relax Nashua, NH: Athena Scientific, 2015.

\bibitem{BeckTeb:09}
A.~Beck and M.~Teboulle, ``A fast iterative shrinkage-thresholding algorithm
  for linear inverse problems,'' \emph{SIAM journal on imaging sciences},
  vol.~2, no.~1, pp. 183--202, 2009.

\end{thebibliography}

\end{document}